\documentclass[preprint,showpacs,preprintnumbers,amsmath,amssymb]{revtex4}

\usepackage{graphicx}
\usepackage{dcolumn}
\usepackage{bm}

\begin{document}

\preprint{}

\title{Silver mean conjectures for 15-$d$ volumes and 
14-$d$ hyperareas
of the separable two-qubit systems}

\author{Paul B. Slater}%
\email{slater@kitp.ucsb.edu}
\affiliation{%
ISBER, University of California, Santa Barbara, CA 93106\\
}%
\date{\today}

\begin{abstract}
Extensive numerical integration results lead us
to conjecture that the {\it silver mean}, that is, 
$\sigma_{Ag} = \sqrt{2}-1 \approx .414214$
plays a fundamental role in certain geometries (those given by monotone metrics) imposable on the {\it 15}-dimensional
convex set of two-qubit systems. For example, we hypothesize  that the volume
of separable two-qubit states, as measured in terms of (four times) the
minimal monotone or Bures metric is $\frac{\sigma_{Ag}}{3}$, and $10 
\sigma_{Ag}$ in terms of (four times) the Kubo-Mori monotone 
metric.
Also, we conjecture, in terms of (four times) the Bures metric,
 that that part of the {\it 14}-dimensional boundary
of separable states consisting generically of rank-{\it four} $4 \times 4$ 
density matrices has  volume (``hyperarea'')
$\frac{55 \sigma_{Ag}}{39}$,
 and that part composed of rank-{\it three} density matrices,
$\frac{43  \sigma_{Ag}}{39}$, so the {\it total} boundary hyperarea would be
$\frac{98 \sigma_{Ag}}{39}$.
 While the Bures {\it probability} of
separability ($\approx 0.07334$)
dominates that ($\approx 0.050339$) based on the Wigner-Yanase metric (and all
other monotone metrics) for
rank-four states, the Wigner-Yanase ($\approx 0.18228$) 
strongly dominates the
Bures ($\approx 0.03982$) for the rank-three states.
\end{abstract}

\maketitle

\section{Introduction} \label{Introduction}
\subsection{Background} \label{Background}
An arbitrary state of two quantum bits (qubits) is describable by a $4 \times 4$ density
matrix ($D_{4}$) --- an Hermitian, 
nonnegative definite matrix having trace unity.
 The convex set of all such density matrices is {\it 15}-dimensional in
nature \cite{vanik,fano}. Endowing this set with the statistical
distinguishability (SD) metric \cite{sam} (identically {\it 
four} times the Bures [minimal monotone]
metric \cite{sam}), we addressed in \cite{slaterqip} 
 the question (first essentially raised in the pioneering study 
\cite{ZHSL}, and 
investigated further in \cite{zycz2,slaterA,slaterC}) 
of what proportion of the 15-dimensional convex
set (now a Riemannian manifold) is separable (classically correlated) in
nature
\cite{werner}. This pertains to the 
question of manifest interest 
``Is the world more classical or more quantum?'' \cite{ZHSL}.

The Peres-Horodecki partial transposition criterion 
\cite{asher,michal}  provides
a convenient necessary {\it and} sufficient condition for testing for 
separability in the cases of
qubit-qubit (as well as qubit-qutrit) pairs \cite{qq}.
That is, if one transposes in place the four $2 \times 2$ blocks of $D_{4}$,
then in the case that  the 
four eigenvalues of the resultant matrix are {\it all} nonnegative --- or
more simply, if its determinant is nonnegative 
\cite[Thm. 5]{ver}--- $D_{4}$ itself is separable.

Sommers and \.Zyczkowski  
\cite[eq. (4.12)]{hans1} have recently established (confirming {\it en passant}
certain conjectures of Slater \cite{slaterhall}) that the
Bures volume of the $(N^2-1)$-dimensional convex set of complex density 
matrices ($D_{N}$) 
 of size $N$ is equal to 
\begin{equation}
\frac{2^{1-N^2} \pi^{N^2/2}}{\Gamma(N^2/2)}.
\end{equation}
For the only specific case of interest here, $N=4$, this gives us for the
total Bures volume,
\begin{equation}
V^{s+n}_{Bures} = \frac{\pi^8}{165150720} \approx 5.74538 \cdot 10^{-5}.
\end{equation}
We
let  the superscript 
$s$ denote the set of separable and the superscript 
$n$ the (complementary) set
of nonseparable $4 \times 4$ density matrices.
(The comparable volume
based
on the {\it Hilbert-Schmidt} metric --- which induces the {\it flat}, 
Euclidean geometry into the set of mixed quantum states --- is 
$V^{s+n}_{HS} = \frac{\pi^6}{851350500}
\approx 1.12925 \cdot 10^{-6}$ \cite[eq. (4.5)]{hans2}.)
The  volume $V^{s+n}_{Bures}$ is {\it exactly}
 equal to that of a 15-dimensional {\it half}sphere
with radius $\frac{1}{2}$ \cite{hans1}.
Now, additionally,
\begin{equation}
\label{eq1}
V^{s+n}_{SD} = 2^{15} V^{s+n}_{Bures} = \frac{\pi^8}{5040} \approx 1.882645.
\end{equation}
So, $V^{s+n}_{SD}$ is itself 
exactly equal to one-half the volume (``surface area'') of a 15-dimensional
sphere of radius
 1.  (The full 
sphere of 
total surface area $\frac{\pi^8}{2520}=2 V^{s+n}_{SD}$ sits in {\it 
16}-dimensional Euclidean
 space and bounds the unit ball there.)

One of the  objectives in this study 
 will be to highly accurately estimate the included volume
$V^s_{SD}$. Then, we could, in turn, 
 obtain a good 
 estimate of the SD/Bures probability of separability.
\begin{equation}
P^s_{SD}= \frac{V^s_{SD}}{V^{s+n}_{SD}} = 
P^s_{Bures} = \frac{V^s_{Bures}}{V^{s+n}_{Bures}}.
\end{equation}
Also, we could
gain evidence as to  possible {\it exact} values, which on the basis of
previous lower-dimensional analyses \cite{slaterC}, 
we have been led to believe is a
distinct possibility.

We had already 
undertaken this task in \cite{slaterqip} (seeking there 
to exploit
the then just-developed Euler angle parameterization of
the $4 \times 4$ density matrices \cite{tbs}). The analysis was, however,
in retrospect, based on a relatively small number (65 million) of points,
generated in the underlying quasi-Monte Carlo procedure (scrambled Halton
sequences) (cf. \cite{lyness,giray1}). (Substantial computer assets were required, nonetheless. Numerical integration in high-dimensional spaces is a particularly
challenging computational task.) One of the classical ``low-discrepancy''
sequences is the {\it van der Corput} sequence in base $b$, where $b$ is any
integer greater than one. The uniformity of the van der Corput numbers can be
further improved by permuting/scrambling
 the coefficients in the digit expansion
of $N$ in base $b$. The scrambled Halton sequence in $N$-dimensions --- which
we employed in \cite{slaterqip} and in our auxiliary analyses below
(sec.~\ref{Auxiliary}) --- is
constructed using
the so-scrambled van der Corput numbers for $b$'s ranging over
the first $N$ prime numbers \cite[p. 53]{giray1}.

To facilitate comparisons with the results of Sommers and \.Zyczkowski
 \cite{hans1}, which were reported subsequent to our
analysis in \cite{slaterqip},
we need to both divide the estimates given 
in \cite{slaterqip} by
$4!=24$ to take into account the strict ordering of the four eigenvalues 
of $D_{4}$ employed by Sommers and \.Zyczkowski
\cite[eq. (3.23)]{hans1},
as well as to multiply them by 8, 
since we (due to a confusion of scaling constants) only, in effect, 
used a factor
of $2^{12}$ in \cite[eq. (5)-(7)]{slaterqip} 
rather than one of $2^{15}$, as indicated above in (\ref{eq1}) 
is required.
These two independent 
adjustments together amount to a  multiplication by $8/24 =1/3$.
This means that 
the estimate of $V^{s+n}_{SD}$ (the true value of which, as given  above, 
is known to be $\approx 1.882645$)  from the
quasi-Monte Carlo analysis in \cite{slaterqip},
should be taken to be 1.88284 = 5.64851/3; 
the estimate of $V^{s}_{SD}$ from \cite{slaterqip} 
should, similarly,  be considered to be 
 0.138767 = .416302/3; and of $P^s_{SD}$ 
(for which no adjustment is needed, being a ratio), 0.0737012.

We had been led in \cite{slaterqip} --- if only for numerical rather
than any clear conceptual reasons --- to formulate a conjecture 
 that (after adjustment by the indicated factor of $\frac{1}{3}$) 
 can be expressed here as
\begin{equation}
\label{eq2}
V^s_{SD} = \frac{\pi^6}{6930} = 0.138729,
\end{equation}
as well as that
\begin{equation}
\label{fundamental}
P^s_{SD} \equiv P^s_{Bures} = \frac{8}{11 \pi^2} \approx 0.0736881
\end{equation}
(suggesting that the [quantum] ``world'' --- even in the case of only
two qubits --- is considerably ``more quantum than
classical''). We now must view (\ref{eq2}) and (\ref{fundamental}) 
as but approximations to
the revised conjectures (\ref{revised}) and  (\ref{eqnew}) below, obtained
on the basis of such much larger quasi-Monte Carlo calculations.

\subsection{Monotone Metrics and Quasi-Monte Carlo Procedures}

The Bures metric plays the role of the {\it minimal}
 monotone metric.
The monotone
metrics comprise an infinite (nondenumerable) class \cite{petz1,petz2,lesniewski}, generalizing the (classically {\it 
unique}) Fisher information metric \cite{kass}.
 The 
{\mbox Bures}
 metric has certainly been the most widely-studied member of this class
\cite{sam,hans1,hubner1,hubner2,ditt1,ditt2}.
For the
 infinitesimal distance element between two states $D_{4}$ and $D_{4} + \delta D_{4}$, we have
\begin{equation}
(ds_{Bures})^2 = \frac{1}{2} \Sigma_{j,k}(\lambda_{j} +\lambda_{k})^{-1} 
| \langle j|\delta D_{4}| k \rangle |^2,
\end{equation}
where $D_{4}$ is diagonal in the orthonormal basis $\{ |j \rangle \}$
with eigenvalues $\{ \lambda_{j} \}$.

Two other prominent members are the {\it maximal} 
monotone metric \cite{yuenlax} and the {\it Kubo-Mori} (KM)
\cite{hasegawa,petz3,michor} 
(also termed Bogoliubov-Kubo-Mori and Chentsov \cite{streater}) 
monotone metric. The Kubo-Mori metric (or canonical correlation stemming from
differentiation of the relative entropy)  is, up to a scale factor, the {\it unique}
monotone Riemannian metric with respect to which the exponential and mixture
connections are {\it dual} \cite{streater}, and as such, certainly 
merits further attention.

In this study, we will utilize additional computer power 
recently available to us,
together with {\it another} advanced quasi-Monte Carlo procedure 
(scrambled Faure-Tezuka sequences \cite{tezuka} --- the use of which was recommended to us
by G. \"Okten, who  provided a corresponding MATHEMATICA 
code). Faure and Tezuka were guided ``by the construction $C^{(i)} = A^{(i)} 
P^{(i-1)}$ and by some possible extensions of the generator formal series in
the framework of Neiderreiter''. ($A^{(i)}$ is an arbitrary nonsingular lower
triangular [NLT] matrix, $P$ is the Pascal matrix \cite{call} 
and $C^{(i)}$ is a generator matrix of a sequence $X$). 
Their idea was to multiply from the right by 
nonsingular upper triangular (NUT) random matrices and get the new generator
matrices $C^{(i)}= P^{(i-1)} U^{(i)}$ for $(0,s)$-sequences 
\cite{tezuka}.``Faure-Tezuka scrambling 
scrambles the digits of $i$ before multiplying by the generator matrices \ldots The effect of the Faure-Tezuka-scrambling can be thought of as reordering
the original sequence, rather than permuting its digits like the Owen
scrambling \ldots Scrambled sequences often have smaller
discrepancies than their nonscrambled counterparts. Moreover, random
scramblings facilitate error estimation'' \cite[p. 107]{hong}.

The Faure-Tezuka procedure appears to us to be 
exceptionally successful in generating a highly uniform (low discrepancy
 \cite{matou}) 
distribution of points over the hypercube --- as judged by its
yielding an estimate of 1.88264 for $V^{s+n}_{SD} \approx 1.882645$.
 However, at this stage,
the procedure {\it does} have the arguable 
shortcoming
 that it does not readily lend itself to the use of
``error bars'' for the estimates it produces, as 
quite naturally do (the generally considerably 
less efficient)
Monte Carlo methods (which, of course, distribute points on the basis of
pseudorandom, rather than deterministic, methods.)

``It is easier to estimate the error of Monte Carlo methods because one can
perform a number of replications and compute the variance. Clever randomizations of quasi-Monte Carlo methods combine higher accuracy with practical
error estimates'' \cite[p. 95]{hong}.
 G. \"Okten is presently developing a MATHEMATICA version of the scrambled 
Faure-Tezuka sequence in which
there will be 
 a random generating matrix for each dimension --- rather than
one for all [fifteen] dimensions --- which will then be
susceptible to statistical testing \cite{hong}. 

\subsection{Morozova-Chentsov Functions} \label{MCF}
To study such monotone metrics {\it other} than the SD/Bures one, we will
utilize a certain {\it ansatz} (cf. \cite{slaterA}).
Contained in the formula \cite[eq. (3.18)]{hans1} of Sommers and \.Zyczkowski for the ``Bures
volume of the set of mixed quantum states'' 
is the subexpression (following their notation),
\begin{equation} \label{ue}
Q_{N}
= \Pi_{\nu < \mu}^{1 \ldots N} \frac{(\rho_{\nu} -\rho_{\mu})^2}{\rho_{\nu} + \rho_{\mu}},
\end{equation}
where $\rho_{\mu},\rho_{\nu}$ ($\mu,\nu=1,\ldots,N$) denote the eigenvalues
of an $N \times N$ density matrix ($D_{N}$).
The term (\ref{ue}) can equivalently be rewritten using the ``Morozova-Chentsov'' function for the Bures metric \cite[eq. (2.18)]{hans1},
\begin{equation} \label{rrr}
 c_{Bures}(\rho_{\mu},\rho_{\nu}) = \frac{2}{\rho_{\nu}
 + \rho_{\mu}},
\end{equation}
as
\begin{equation} \label{xxx}
Q_{N}= \Pi_{\nu < \mu}^{1 \ldots N} (\rho_{\nu}
-\rho_{\mu})^2 c_{Bures}(\rho_{\mu},\rho_{\nu})/2.
\end{equation} 
A Morozova-Chentsov function is a positive continuous function $c(\lambda,\mu)$ that is symmetric in its two variables and for which $c(\lambda,\lambda)
= C \lambda^{-1}$, for some constant $C$, and $c(t \lambda, t \mu) = t^{-1}
c(\lambda,\mu)$ \cite[Thm. 1.1]{petz2}. There exist one-to-one correspondences
 between Morozova-Chentsov functions, monotone metrics and operator means.
\cite[Cor, 6]{petz2}. ``Operator means are binary operations on positive operators which fulfill the main requirements of monotonicity and the transformer
inequality'' \cite{petz2}.

The {\it ansatz} we employ 
is that the replacement of $c_{Bures}(\rho_{\mu},\rho_{\nu})$
 in the formulas for the Bures volume 
element
 by the particular
Morozova-Chentsov
function corresponding to a given monotone metric ($g$) will
yield the
volume element
 corresponding to that particular 
$g$. We have been readily able to
validate this for a number of instances
 in the case of the {\it two}-level quantum systems  
[$N=2$], using
the general formula for the monotone metrics over 
such systems of Petz and Sud\mbox{\'a}r \cite[eq. (3.17)]{petz1}.
One can argue
that the joint distribution of the eigenvalues 
of $D_{N}$ 
is the product
of $Q_{N}$ --- pertaining to the off-diagonal elements of the density
matrix ---- and an additional factor $H_{N}$--- pertaining to the diagonal
elements. Now, $H_{N}$
is equal to the reciprocal of the square root of
the determinant of the density matrix 
for {\it all} [Fisher-adjusted] 
monotone metrics --- so we need not be concerned with its variation
across metrics in this study ---  and simply unity in the case of the
[flat] Hilbert-Schmidt metric (cf. \cite{hall}).
\subsection{Outline of the study}
In addition to studying the SD/Bures metric, 
we ask analogous questions in relation to 
a number of  other monotone metrics of interest. We study two of these
 metrics, in addition to the SD metric, 
 in our ``main analysis'' (sec.~\ref{main}) and two more
 in our ``auxiliary analysis'' (sec.~\ref{Auxiliary}),
 which is based on the same 
scrambled Halton procedure employed in \cite{slaterqip} --- but with more
than five times the number of points generated there, but also many
fewer points than in the primary (main) analysis here.
(In hindsight, we might have better
 consolidated the several 
 monotone metrics into a {\it single}
investigation, from the very outset, 
but our initial/tentative/exploratory analyses grew, and we 
were highly reluctant to discard several
weeks worth of demanding and apparently 
revealing 
computations. Also, we had been using two different sets of processors
[Macs and Suns]
for our computations and for a number of reasons --- too involved and
idiosyncratic to make the subject of discussion here --- it proved convenient
to conduct {\it two} distinct analyses.) Also, we include analyses
in sec.~\ref{Maximal}  pertaining to the {\it maximal} monotone metric, and a number
of metrics interpolated between the minimal and maximal ones.
(The ``average'' monotone 
metric --- studied in our main analysis (sec.~\ref{main}) --- is obtained by such an interpolation.) In sec.~\ref{MonteCarlo} we apply Monte-Carlo
methods in a limited study of the questions raised before.
 In sec.~\ref{Boundary}, we undertake studies concerned with
the values of volumes (``surface areas'') 
of the 14-dimensional {\it boundary} of the
15-dimensional convex set of two-qubit states, as measured in terms of
the various monotone metrics under investigation here.

To  begin with (sec.~\ref{Preliminary}), we
 will  seek 
to  determine  $V_{\tilde{KM}}^{s+n}$.
 A wiggly line over the acronym for a metric will denote that we have 
{\it ab initio} multiplied that
metric by 4, in order to facilitate comparisons with results 
presented in terms of the SD, rather than the
Bures metric, which is one-fourth of the SD metric. 
(This, perhaps fortuitously, gives us a quite appealing {\it scale} 
of numerical results.) The probabilities themselves --- being computed as {\it ratios} --- are,
 of course, invariant under such a scaling, so the ``wiggle'' is irrelevant
for them.
\section{Preliminary Analysis of the Kubo-Mori Metric} \label{Preliminary}
The Morozova-Chentsov function for the Kubo-Mori metric is \cite[eq. (2.18)]{hans1}
\begin{equation} \label{genform}
c_{KM}(\rho_{\mu},\rho_{\nu}) = 
\frac{\log{\rho_{\nu}} -\log{\rho_{\mu}}}{\rho_{\nu} - \rho_{\mu}}.
\end{equation}
To proceed in the study of the KM metric, we first wrote a 
MATHEMATICA program, using the numerical integration command, 
 that succeeded to a high
degree of accuracy in reproducing the formula \cite[eq. (4.11)]{hans1},
\begin{equation} \label{hall}
C_{N} =  \frac{2^{N^2-N} \Gamma (N^2/2)}{\pi^{N/2} \Gamma (1)  \ldots 
\Gamma (N+1)}
\end{equation} 
for the Hall/Bures
 normalization constants \cite{hall,slaterhall} for various $N$. (These
constants
 form one of the two factors --- along with the volume of the
flag manifold \cite[eqs. (3.22), (3.23)]{hans1} --- in determining the total Bures volume.)
Then, in the MATHEMATICA program,
 we replaced the Morozova-Chentsov function (\ref{rrr}) for the Bures metric
 in the product formula 
(\ref{xxx})
by the one (\ref{genform}) 
for the Kubo-Mori function. For the cases $N=3,4$ we found
that the new numerical results were to several decimal places of accuracy (and in the case $N=2$, exactly)
 equal to $2^{N(N-1)/2}$ times the comparable result for the Bures
 metric, given by (\ref{hall}). This immediately implies that the KM volumes
of mixed states are also $2^{N (N-1)/2}$ times the corresponding Bures 
volumes (and the same for the $\tilde{KM}$ and SD volumes),
 since the remaining factors involved, that is, the volumes of the flag
manifolds are common to both the Bures and KM cases (as well as to all
the monotone metrics).
Thus, we arrive at our first 
{\it conjecture} (cf. \cite{hans1}),
\begin{equation} \label{mmm}
V_{\tilde{KM}}^{s+n} = 64 V_{SD}^{s+n} = \frac{4 \pi^8}{315} \approx 120.489,
\end{equation}
for which we will obtain some 
further support in our {\it main} 
numerical analysis (sec.~\ref{main}),  yielding
Table~\ref{tab:table1}.
\begin{table}
\caption{\label{tab:table1}Estimates based
on (four times) 
the Bures, ``average''  and Kubo-Mori metrics, using a scrambled
Faure-Tezuka sequence composed of {\it two} billion points distributed over the
15-dimensional unit hypercube,  for 
quasi-Monte Carlo numerical integration. The results based on 
the first {\it one}
billion points
are given in parentheses.}
\begin{ruledtabular}
\begin{tabular}{rccc}
metric & $V^{s+n}_{\tilde{metric}}$ & $V^{s}_{\tilde{metric}}$ & 
$P_{metric}^s = V^s/V^{s+n}$ \\
\hline
Bures & 1.88264 (1.88264) & 0.137884 (0.137817)  & 0.0732398  (0.0732042) \\
Average & 28.0801 (28.0803)   & 1.33504 (1.33436)  & 0.0475438  (0.0475194) \\
Kubo-Mori & 120.504 (120.531)  & 4.1412 (4.14123)  & 0.0343654  (0.0343583)\\
\end{tabular}
\end{ruledtabular}
\end{table}

\section{Main Analysis} \label{main}

Associated with the minimal (Bures) monotone metric is the operator monotone function, $f_{Bures}(t) = (1+t)/2$, and with
the {\it maximal} monotone metric, the operator monotone function, $f_{max}(t)=2 t/(1+t)$ \cite[eq. (2.17)]{hans1}.
The {\it average}
 of these two functions, that is, $f_{average}(t)=(1+6 t + t^2)/(4 +4 t)$, 
is also necessarily 
operator monotone \cite[eq. (20)]{petz2} and thus 
 yields 
 a monotone metric (apparently previously uninvestigated). Again employing our basic {\it ansatz}, we used the associated Morozova-Chentsov function --- given by the general formula \cite[p. 2667]{petz1},
$c(x,y)=1/y f(x/y)$ ---
\begin{equation}
c_{average}(\rho_{\mu},\rho_{\nu}) = \frac{4 (\rho_{\mu} +\rho_{\nu})}{\rho_{\mu}^2 + 6 \rho_{\mu} \rho_{\nu} + \rho_{\nu}^2}.
\end{equation}

For our main quasi-Monte Carlo 
analysis, we (simultaneously) numerically 
integrated the SD, $\tilde{KM}$ and $\tilde{avg}$
 volume elements
 over a fifteen-dimensional {\it hypercube} using two {\it billion} 
points for evaluation, with the  points forming a  scrambled {\it Faure-Tezuka}
sequence \cite{tezuka}.
 (As in \cite{slaterqip}, the fifteen original variables --- {\it twelve} Euler angles and {\it three}
 angles for the 
eigenvalues  \cite[eq. (38)]{tbs} --- parameterizing
 the $4 \times 4$ density matrices
were first linearly transformed so as to all lie in the range
[0,1].) This ``low-discrepancy''
sequence is designed to give a close-to-{\it uniform} coverage
of points over the hypercube, and accordingly yield 
relatively accurate numerical
integration results.

The results of Table~\ref{tab:table1} suggest to us, now, rejecting 
the previous conjecture (\ref{eq2}) --- based on a much smaller
number (65 million) of data points than the two billion here --- and 
replacing it 
by (perhaps the more ``elegant'')
\begin{equation} 
\label{revised}
V^s_{SD} = \frac{\sigma_{Ag}}{3} \equiv  \frac{\sqrt{2}-1}{3}
 \approx 0.138071,
\end{equation}
where $\sigma_{Ag}$ denotes the ``silver mean'' \cite{christos}.
(As we proceed from one billion to two billion points, some 
apparent convergence --- 0.137817 to 0.137884 --- of the numerical estimate 
to the conjecture  (\ref{revised}) is observed.
It is interesting to note the occurrence of the {\it first}
 three positive
integers in (\ref{revised}) --- a property which obviously the much-studied
{\it golden mean}, $\frac{\sqrt{5}-1}{2}$ lacks.) 
 By
implication then, the conjecture (\ref{fundamental}) is replaced by
\begin{equation}
\label{eqnew}
P^s_{SD/Bures}= \frac{V^s_{SD}}{V^{s+n}_{SD}} = 
\frac{1680 \sigma_{Ag}}{\pi^8}
 \approx 0.0733389.
\end{equation}

In addition to simply our numerical results, we were also encouraged to advance this conjecture (\ref{revised})
on the basis of certain earlier results. In  \cite{slaterC}, a number of quite 
surprisingly simple {\it 
exact} results
were obtained using {\it symbolic} integration, for certain specialized 
[low-dimensional] 
two-qubit scenarios. This had led us to first investigate
in \cite{slaterqip} the possibility of an
exact probability of separability also 
in the {\it full} 15-dimensional setting. (Unfortunately, as far as we can perceive, the
full 15-dimensional 
problem is not at all amenable --- due to its complexity --- to the use of the 
currently available symbolic
integration programs and, it would appear, possibly for the foreseeable
future.)

In particular in
 \cite{slaterC}, a Bures {\it probability} of separability equal to
$\sigma_{Ag}$ had been obtained
 for both the $q=1$ and $q = \frac{1}{2}$ states \cite{abe} 
inferred using the principle of maximum nonadditive [Tsallis] 
entropy --- and also for an additional {\it low}-dimensional scenario \cite[sec. II.B.1]{slaterC}.
(We have recently reanalyzed this last scenario, but with the {\it maximal}
monotone metric, and also found a probability of separability equal to
$\sigma_{Ag}$.
The value $\sigma_{Ag}$ also arises as the amount by which Bell's inequality
is violated \cite[eq, (8)]{gill}.) Christos and Gherghetta \cite{christos}
took the silver mean, as in our study here,
 to be the positive solution of the equation 
$x+2 =\frac{1}{x}$ --- since having a ``mean'' value less than 1 was useful
in their investigation
 of trajectory scaling functions --- while others (perhaps more) 
 \cite{spinadel,gumbs} \cite[chap. 22]{kappraff}
 have defined it as the positive
solution of $x-2=\frac{1}{x}$, that is, $\sqrt{2}+1$, the {\it reciprocal}
of our $\sigma_{Ag}$. (The square root of two minus one 
 is also apparently a form of
``Pisot number'' \cite{escudero}. Similar definitional, but perhaps 
not highly signficant,
 ambiguities occur in the (more widespread) usage of the term
``golden mean'', that is $(\sqrt{5} \pm 1)/2$ \cite{carlosfriend}.
(``The characteristic sequence of $(\sqrt{5}-1)/2$ (resp., $\sqrt{2}-1$)
 is called
the {\it golden mean sequence} (resp., {\it Pell sequence})'' \cite{chuan}.
This line of analysis --- concerned with the alignment of two words
over an alphabet --- originated, apparently, from 
 a 1963 unpublished talk of the
prominent [Nobelist] physicist, D. R. Hofstadter \cite{chuan}.)
In \cite{bayard}, demonstrating a conjecture of Gromov, 
the {\it minimal volume} of $\bf{R}^2$ (the infinite Euclidean plane) 
was shown
to be $\frac{2 \pi}{\sigma_{Ag}}$. (An exposition of this result is given
in \cite{bowditch}.) In \cite{bambah} the value of $\frac{1}{2 \sigma_{Ag}}$ 
 was obtained for a certain supremum of {\it volumes}.

Further  conjectures that $V^s_{\tilde{avg}}= \frac{29 \sigma_{Ag}}{9} 
\approx 1.33469$ and 
$V^s_{\tilde{KM}} = 10 \sigma_{Ag} \approx 4.14214$ seem 
 worth investigating, based
on the results in Table~\ref{tab:table1}.
(Our estimate of the ratio $\frac{V^{s}_{\tilde{KM}}}{V^{s}_{SD}}$ 
from Table~\ref{tab:table1} is 30.0339.)
So, we have an implied conjecture that 
\begin{equation} \label{MK}
P^s_{KM} =\frac{V^s_{\tilde{KM}}}{V^{s+n}_{\tilde{KM}}} = \frac{1575 
\sigma_{Ag}}{2 \pi^8} \approx .0343776.
\end{equation}
It would then follow that 
 $\frac{P^s_{KM}}{P^s_{SD/Bures}}=\frac{15}{32}=.46875$.

The convergence to
 the {\it known}
 value of $V^{s+n}_{SD}$ in
 Table~\ref{tab:table1} seems more pronounced than any presumptive 
convergence to the conjectured
values of the separable volumes alone, but the latter are
based on considerably smaller samples 
(roughly, one-quarter the number) of points than the former (for which, of
course, {\it all} the two billion systematically generated
points are used).
Clearly, our conjecture (\ref{revised}) can be reexpressed as
$V^n_{SD} = V^{s+n}_{SD}- V^s_{SD}=
 \frac{\pi^8}{5040} - \frac{\sigma_{Ag}}{3} \approx 1.74457$.
 Our sample estimate
for $V^n_{SD}$ is, then, 1.74475.)

\begin{figure}
\includegraphics{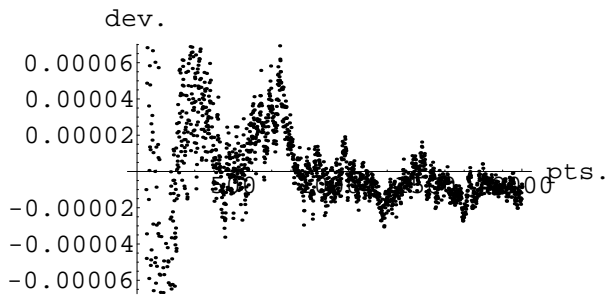}
\caption{\label{fig:SDsepnon} Deviations of the estimated  values of
$V^{s+n}_{SD}$ from the {\it known}  value --- as shown by
Sommers and \.Zyczkowski --- of $\frac{\pi^8}{5040} 
\approx 1.882645$, as the number of points
in the scrambled Faure-Tezuka sequence increases from 1 million
to 2,000 million}
\end{figure}

\begin{figure}
\includegraphics{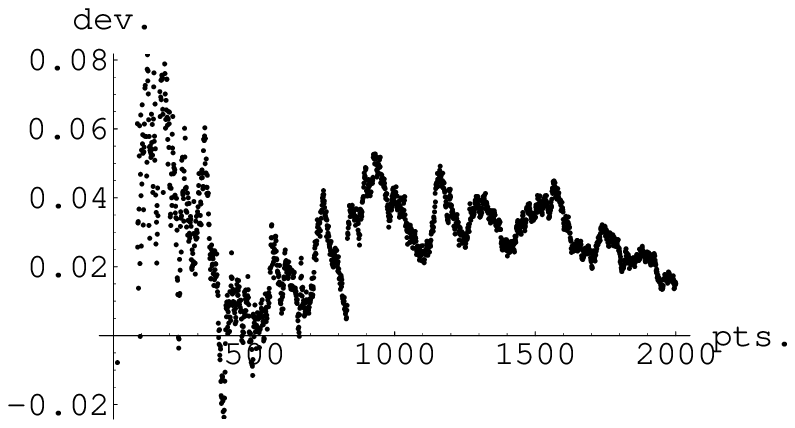}
\caption{\label{fig:KMsepnon} Deviations of the estimated  values of
$V^{s+n}_{\tilde{KM}}$ from the 
{\it conjectured} value of $\frac{4 \pi^8}{315} 
\approx 120.489$, as the number of points
in the scrambled Faure-Tezuka sequence increases from 1 million
to 2,000 million}
\end{figure}

\begin{figure}
\includegraphics{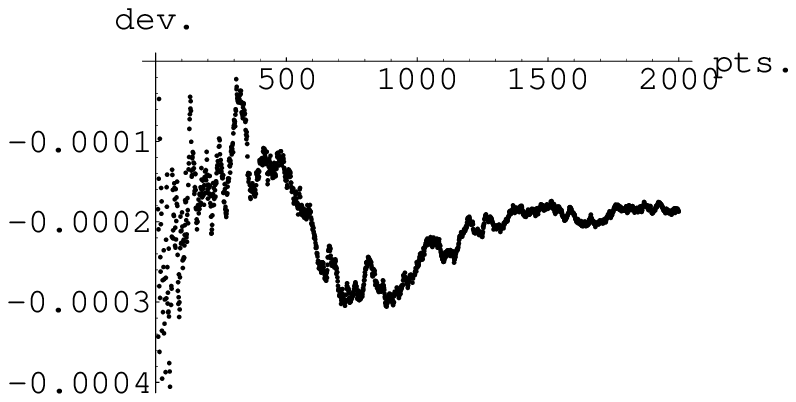}
\caption{\label{fig:SDsep} Deviations of the estimated  values of
$V^s_{SD}$ from the {\it conjectured} value of $\frac{\sigma_{Ag}}{3} \approx 
0.138729$, as the number of points
in the scrambled Faure-Tezuka sequence increases from 1 million
to 2,000 million}
\end{figure}

\begin{figure}
\includegraphics{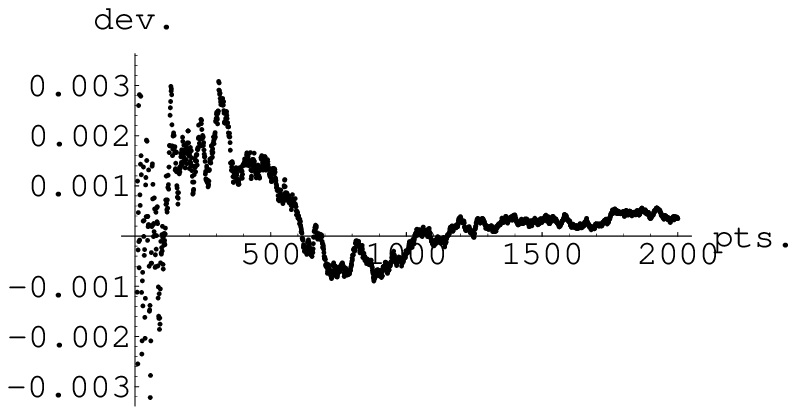}
\caption{\label{fig:averagesep} Deviations of the estimated values of
$V^s_{\tilde{avg}}$ from the {\it conjectured} value of
$\frac{29 \sigma_{Ag}}{9} \approx  1.33469$, as the number of points
in the scrambled Faure-Tezuka sequence increases from 1 million
to 2,000 million}
\end{figure}

\begin{figure}
\includegraphics{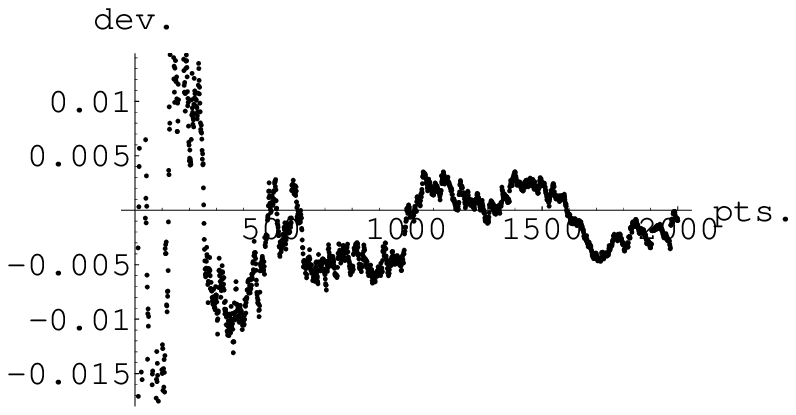}
\caption{\label{fig:KMsep} Deviations of the estimated values of
$V^s_{\tilde{KM}}$ from the {\it conjectured} value of $10 \sigma_{Ag} 
\approx  4.14214$,
 as the number of points
in the scrambled Faure-Tezuka sequence increases from 1 million
to 2,000 million}
\end{figure}

In Figs.~\ref{fig:SDsepnon}-\ref{fig:KMsep},
 we show the {\it deviations} from our conjectured and known
values of the estimates provided by the Faure-Tezuka sequence as the
number of points in the sequence 
increases from one million to {\it two thousand}
 million
(i. e. two {\it billion}).

\begin{figure}
\includegraphics{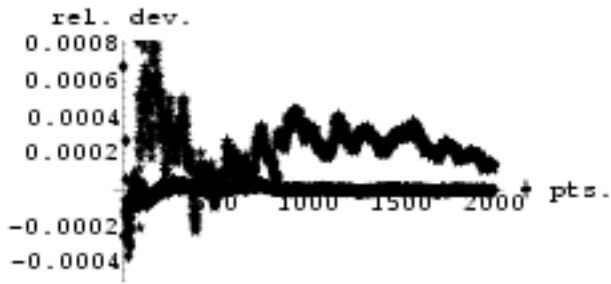}
\caption{\label{fig:reldev} Joint plot of {\it relative}
 deviations of estimates
of $V^{s+n}_{SD}$ and $V^{s+n}_{\tilde{KM}}$ from their known and 
conjectured values
of $\frac{\pi^8}{5040}$ and $\frac{ 4 \pi^8}{315}$. The more rugged
curve corresponds to $V^{s+n}_{\tilde{KM}}$.}
\end{figure}

In Fig.~\ref{fig:reldev}, additionally, we show {\it together} 
the {\it relative} deviations of
$V^{s+n}_{SD}$ and of $V^{s+n}_{\tilde{KM}}$ from the known and
conjectured values.
In other words, we {\it divide} the estimated values by the known/conjectured values
and subtract 1. The SD curve is extraordinarily better behaved 
(``hugging'' the $x$-axis) than is the KM curve.
Perhaps this difference is attributable to the ``simpler'' 
(more numerically stable?) nature of the
Morozova-Chentsov function in the SD case (\ref{rrr}) than in the KM
case (\ref{genform}).

\begin{figure}
\includegraphics{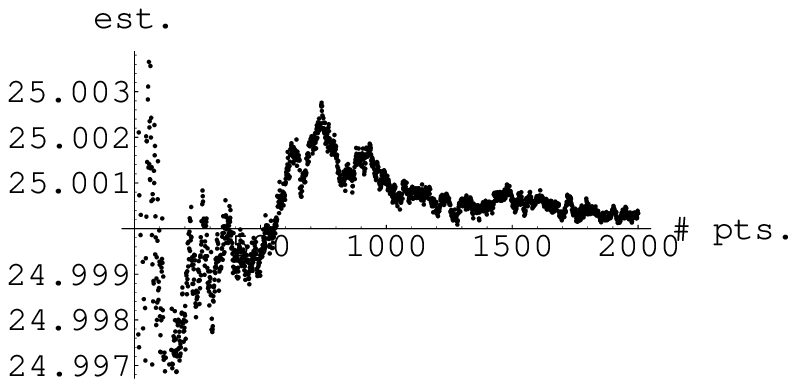}
\caption{\label{fig:SDavg} Estimates of the difference, $V^n_{\tilde{avg}}-V^n_{SD}$,
as the number of points in the scrambled Faure-Tezuka sequence increases
from 1  million to 2,000 million. Note the strong suggestion that the
true value is 25 (or close thereto)}
\end{figure}

Further plotting of our various results yielded one of particular interest.
In Fig.~\ref{fig:SDavg}
 we show the estimates of $V^n_{\tilde{avg}}-V^n_{SD}$.
Of course, this figure (the scale of which was internally chosen by MATHEMATICA, based on the data, and not exogeneously imposed) 
strongly suggests that $V^n_{\tilde{avg}}-V^n_{SD}
= 25$. Now, we found that if we posit
\begin{equation} \label{latest}
V^{s+n}_{\tilde{avg}} =\frac{25 \pi^8}{8448} \approx 28.0792,
\end{equation}
(with $8448 = 2^8 \cdot 3 \cdot 11$), it would follow that
\begin{equation}
V^n_{\tilde{avg}}-V^n_{SD} = \frac{2449 \pi^8}{887040} -\frac{26}{9}
\sigma_{Ag} \approx 24.99996094,
\end{equation}
(with $887040 = 2^8 \cdot 3^2 \cdot 5 \cdot 7 \cdot 11$), being
{\it strikingly} close to the indicated value of 25.

\begin{figure}
\includegraphics{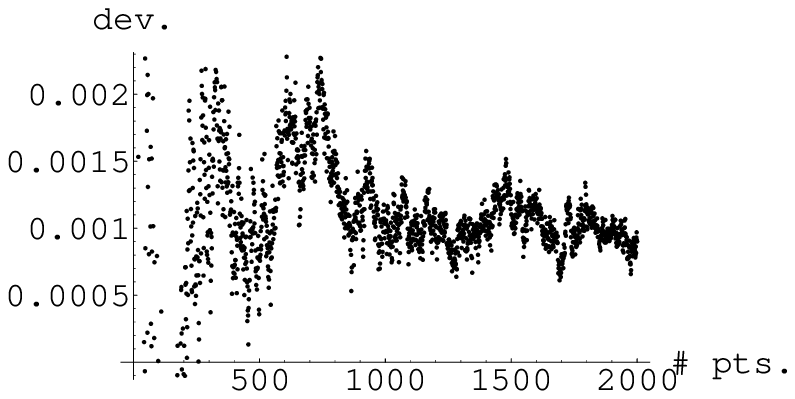}
\caption{\label{fig:avgavg} Deviations of the estimated values 
of $V^{s+n}_{\tilde{avg}}$ 
from the conjectured value of $\frac{25 \pi^8}{8448}$,
as the number of points
in the scrambled Faure-Tezuka sequence increases from 1 million to
2,000 million}
\end{figure}

In Fig.~\ref{fig:avgavg} 
 we plot the deviations of the estimates of $V^{s+n}_{\tilde{avg}}$ from its conjectured value (\ref{latest}).

\section{Auxiliary analysis}  \label{Auxiliary}
In an {\it independent} set of computations (Table~\ref{tab:table2}),
employing 415 million points
of a scrambled {\it Halton} sequence (as opposed to the scrambled Faure-Tezuka
sequence used in sec.~\ref{main}), we sought to obtain estimates of the
probability of separability of two arbitrarily coupled qubits based
on three monotone metrics of interest. The specific method employed was that of scrambled
Halton sequences \cite{giray1}. (While there are different scrambled
Faure-Tezuka sequences depending upon the particular random generating
matrices used, the scrambled Halton sequence is unique in nature.)

These correspond to the operator monotone functions,
\begin{eqnarray}
 f_{GKS}(t)= t^{t/(t-1)}/e; \quad 
f_{WY}(t) = \frac{1}{4} (\sqrt{t}+1)^2; \quad
f_{KM}(t) = \frac{t-1}{\log{t}}.
\end{eqnarray}
The subscript GKS denotes  the Grosse-Krattenthaler-Slater (``quasi-Bures'') 
metric
(which yields the common asymptotic minimax and maximin redundancies
for {\it universal} quantum coding \cite[sec. IV.B]{KS} \cite{gillmassar}),
the subscript WY,
the Wigner-Yanase information metric
\cite[sec. 4]{gi} \cite{wy,luo}, and the subscript KM, the Kubo-Mori metric
already studied in secs. II and III. (We had, in fact, intended to study
the ``Noninformative'' monotone metric \cite{slaterclarke} here instead
of the KM metric, but there was a programming oversight that was only uncovered
at the end of the computations.)
\begin{table} 
\caption{\label{tab:table2}Estimates based on
(four times) the Grosse-Krattenthaler-Slater, Wigner-Yanase  and Kubo-Mori 
monotone
metrics, using a scrambled Halton sequence 
consisting of 415 million points.}
\begin{ruledtabular}
\begin{tabular}{rccc}
metric & $V^{s+n}_{\tilde{metric}}$  & $V^{s}_{\tilde{metric}}$ &
 $P^s_{metric} = V^s/V^{s+n}$ \\
\hline
GKS  & 5.4237 & 0.330827  & .0609965 \\
WY & 14.5129 & 0.730567& .0503391 \\
KM  & 120.504 & 4.1791 & .0346801 \\
\end{tabular}
\end{ruledtabular}
\end{table}
It appears conjecturable that, in terms of the separable states, 
 $V^s_{\tilde{GKS}} =\frac{4 \sigma_{Ag}}{5} 
\approx 0.331371$.
(The evidence is somewhat of a weaker nature that $V^s_{\tilde{WY}} 
=\frac{7 \sigma_{Ag}}{4} \approx 0.724874$.)
Also, in terms
of the combined separable and nonseparable states, it seems possible
that $V^{s+n}_{\tilde{GKS}}= \frac{\pi^8}{1750} \approx 5.42202$, with
$1750 = 2 \cdot 5^3 \cdot 7$.
If so, we would have 
$P^s_{GKS} = \frac{1400 \sigma_{Ag}}{\pi^8} \approx 0.0611158$.

We had hoped to further extend the scrambled Halton sequence used here,
but doing so has so far proved problematical, in terms of available
computer resources.
\section{Maximal Monotone Metric} \label{Maximal}
As to the {\it maximal}
 monotone metric, numerical, together with some analytical evidence, strongly indicate that
$V_{max}^{s+n}$ is infinite (unbounded) (as well as $V^s_{max}$).
The supporting 
{\it analytical} evidence consists in the fact 
that for the three-dimensional convex set of $2 \times 2$ density matrices,
parameterized by spherical coordinates [$r,\theta,\phi$] in the ``Bloch
ball'', the volume element of the maximal monotone metric is
$r^2 \sin{\theta} (1-r^2)^{-{3/2}}$, the integral of which {\it 
diverges} over
the  ball. Contrastingly, the volume element of the minimal monotone metric is
$r^2 \sin{\theta} (1-r^2)^{-{1/2}}$, the integral over the ball of which is
{\it finite}, namely
$\pi^2$. For $s \geq 1$ the integral of $r^2 \sin{\theta} (1-r^2)^{-s}$ diverges, so
the divergence associated with the monotone metric itself 
is not simply
 marginal or ``borderline'' 
in character.

To gain further evidence in these regards,
one can engage in 
numerical estimation for the one-parameter 
family of {\it interpolating} metrics given
by the operator monotone functions
\begin{equation}
f_a(t) = (1-a) f_{max}(t) + a f_{Bures}(t),
\end{equation}
for which the Morozova-Chentsov functions are of the form
\begin{equation}
c_{a}(\rho_{\mu},\rho_{\nu}) = \frac{ 2 (\rho_{\mu} + \rho_{\nu})}{a (\rho_{\mu}-\rho_{\nu})^2 + 4 \rho_{\mu} \rho_{\nu}}.
\end{equation}
Then, one could 
plot 
the results as a function of the parameter $a$ and study the
limit $a \rightarrow 0$. (Of course, for $a =\frac{1}{2}$, 
one would recover the ``average'' monotone metric, studied in sec.~\ref{main}.)

A preliminary investigation along these lines is reported in 
Table~\ref{tab:table3}. 
\begin{table}
\caption{\label{tab:table3}Estimates based on the first 
ninety-six million points
of a scrambled Faure-Tezuka sequence of  (four times) a number of metrics obtained
by
interpolating between the maximal ($a=0$) and minimal ($a=1$) monotone
metrics}
\begin{ruledtabular}
\begin{tabular}{cccccccc}
a & 1  & $10^{-1}$ & $10^{-2}$ & $10^{-3}$ & $10^{-4}$  & $10^{-5}$ & 0 \\
\hline
$V^{s+n}_{a}$ & 1.88258 & 7951.27 & $9.3254 \cdot 10^6$ & $6.0345 \cdot 10^9$ & $3.049 \cdot 10^{12}$ &  $1.2825 \cdot 10^{15}$ & $ 8.0858 \cdot 10^{39}$  \\
$V^s_{a}$ & 0.13786 & 148.569 & 63659. & $2.0972 \cdot 10^7 $ & 
$6.502 \cdot 10^{9}$ & $ 2.2084 \cdot 10^{12}$ & $5.229 \cdot 10^{36}$ \\
$P^s_{a}$ & 0.073229 & 0.01868 & 0.006827 & 0.003475 & 0.0021325 & 0.0017219 & 0.00064669  \\
\end{tabular}
\end{ruledtabular}
\end{table}
 Based on the first {\it ninety-six} 
 million points of a
scrambled Faure-Tezuka sequence, we obtain estimates of $V^{s+n}_{a}$, $V^s_{a}$ 
and $P^s_{a}= \frac{V^s_{a}}{V^{s+n}_{a}}$ 
for (four times) the metrics interpolating between the maximal ($a=0$) and
minimal ($a=1$) monotone metrics for several values of $a$, increasingly 
close to $a=0$.  So, $P^s_{max}$ would seem quite close to being 0.
(However, some clearly numerically anomalous behavior occurred
in passing from ninety-six million points to ninety-seven 
million points. The estimates of $V^{s+n}_{max},
V^s_{max}$ and $P^s_{max}$ jumped to $8.27999 \cdot 10^{40},
7.48026 \cdot 10^{40}$ and $0.903414$, respectively.)

 It would be interesting to {\it formally} test the hypothesis
that $P^s_{max}=0$. (More specifically, we might ask the question if
the {\it limit} of $P^s_{a}$ as $a \rightarrow 0$ is 0.)
If it can, in fact, be established that $P^s_{max}$ is zero, this might serve as something in the
nature of a ``counterexample'' to the proposition (a matter of considerable
interest in the theoretical analysis of quantum computation)
that for bipartite quantum systems of finite dimension, there {\it is} a
separable neighborhood of the fully mixed state of finite
 volume \cite{slb,ch,gb,szarek}. (These conclusions
 were obtained with the use of
either the trace or Hilbert-Schmidt metric --- the first of which is monotone, but not Riemannian, while the second is Riemannian, but not 
monotone \cite{hans1}.)

We have conducted a test along these lines. Using a simple 
{\it 
Monte-Carlo} (rather than quasi-Monte Carlo)
scheme, we generated ten sets of ten million points {\it 
randomly} distributed
over the 15-dimensional hypercube. For each of the ten sets, we obtained
estimates of $V^{s+n}_{max}$, $V^s_{max}$ and hence $P^s_{max}$.
Based on the one hundred 
million points the (mean) estimate of $P^s_{max}$ was 
$\mu = 1.77038 \cdot 10^{-7}$
and the standard deviation across the ten samples, $\eta= 
3.692 \cdot 10^{-7}$.
 So, the
value 0 lies {\it less} than one-half (that is, 0.479510) 
standard deviations from $\mu$. 
For a student $t$-distribution with $9=10-1$
degrees of freedom,
forty  percent of the probability
lies outside  0.261 standard deviations from the mean and twenty-five percent
outside 0.703 standard deviations. So there is 
little evidence here
for rejecting a hypothesis that
$P^s_{max}$ equals 0.  For an independent analysis 
based on ten sets of {\it four} million points,
the estimates were roughly comparable, {\it i. e.}, 
$\mu= 2.4196 \cdot 10^{-7}, \eta = 5.09683 \cdot 10^{-7}$.
Also, for ten sets of {\it five} million points, but setting the
interpolation parameter $a$ not to 0
but to .05, there were obtained $\mu=.00438593$ and $\eta=.000229437$, 
with $\frac{\mu}{\eta}= 19.1611$.
So, here one {\it can} decisively reject a hypothesis that the probability of
separability for $a = .05$ is 0.
\section{Further Monte-Carlo Analyses} \label{MonteCarlo}
We also undertook a  series of {\it Monte-Carlo} 
analyses,
incorporating together the GKS, Bures, average, Kubo-Mori, Wigner-Yanase,
maximal and Noninformative (NI) \cite{slaterclarke}
monotone metrics. (The operator monotone function $f(t)$ associated with the
NI metric is $\frac{2 (t-1)^2}{(1+t) \log^2{(t)}}$.) 
We subdivide
the unit hypercube into $3^{15} = 14,348,907$ subhypercubes, pick a 
{\it random} point in each one of these, and then repeat the procedure...
We are now able to report in Table~\ref{tab:table4}
the results of fifteen  iterations of this process.
The {\it central limit theorem} tells us that for
a large enough sample size, the distribution of the sample mean will approach
a normal/Gaussian distribution. This is true for a sample of independent random
variables from any population distribution, so long as the population
has a {\it finite} standard deviation. The population standard deviation
is equal to the standard deviation of the mean times the square root of
the sample size $N$, which in our case is $15 \cdot 3^{15}$.
If one were to use {\it two} standard deviations as a rejection criterion,
then the only one of our conjectures that would be rejected would be that
for $V^s_{\tilde{GKS}}$. (However, the standard deviations in the separable
cases would be approximately four times as large
if we only used the number of points corresponding to separable states, rather than to {\it all} states, as we have done here. This would lead, then, to 
{\it none}
of our conjectures being rejected.)
\begin{table}
\caption{\label{tab:table4} Monte-Carlo analysis based on 
$15 \cdot 3^{15} = 215,233,605$  density matrices. 
The second set of columns correspond 
to $V^s_{\tilde{metric}}$ and the third, to $V^{s+n}_{\tilde{metric}}$. Estimates (est.) of the volumes, standard deviations (s.d.) of these estimates
 and the number of standard deviations
they are away from their conjectured (cj.) 
or known values, as  given 
in Table~\ref{tab:table5}, are presented}
\begin{ruledtabular}
\begin{tabular}{r|rrr|rrr}
metric & est. & s. d.  & $\frac{cj.-est.}{s. d.}$ & est. & s. d. & 
$\frac{cj. -est.}{s. d.}$ \\
\hline
Bures &  0.13800&  0.00021&  0.3354 & 1.88295& 0.00102& -0.2985\\
GKS & 0.32990  & 0.00057& 2.5696 & 5.4232  & 0.00315& -0.3867  \\
WY  & 0.72811 & 0.00152& ---& 14.5084  & 0.00943&---  \\
Avg &  1.3363 & 0.00287& -0.5821 & 28.0781  & 0.01769 & 0.0618 \\
KM  & 4.1574& 0.02242& -0.6792&120.256 & 0.17489& 1.3363\\
NI  & 848.05 & 28.997& --- & 48668.2& 421.712 & --- \\
\end{tabular}
\end{ruledtabular}
\end{table}
The estimate of $P^s_{max}$, obtained in the same Monte-Carlo procedure,  was
 $\frac{4.33981 \cdot 10^{42}}{1.93678  \cdot 10^{52}} = 2.24073 
 \cdot 10^{-10}$.
\section{{\it 14}-Dimensional Boundaries} \label{Boundary}
\subsection{Initial analyses}
In the analyses above, we have been concerned with the volume of the
15-dimensional convex set of $4 \times 4$ density matrices, as measured in
terms of a number of monotone metrics.
We have modified the computer programs involved, so that they would
provide estimates of the volume (``hyperarea'')  of the boundary of this set. 

Our numerical
integrations were conducted over a {\it fourteen}-dimensional hypercube, now
allowing one of the original fifteen 
variables (specifically, the hyperspherical angle designated $\theta_{3}$ in
\cite[eq. (2)]{slaterqip}) to be determined not by the quasi-Monte Carlo procedure, but by the requirement that the
determinant of the partial transpose equal zero. This considerably
increases the computational effort {\it per} point generated.

 Our early estimates, in this regard, 
 were: 0.587532 (SD), 6.25466 ($\tilde{avg}$) and 19.8296
 ($\tilde{KM}$), all three based on the first 2,600,000 
points of a scrambled Faure-Tezuka sequence; 1.47928 ($\tilde{GKS}$),
3.37384 ($\tilde{WY}$) and 19.9277 ($\tilde{KM}$), all three based on the 
first 1,500,000 points of a scrambled Halton sequence; and
0.588816 (a=1), 837.072 
$(a=10^{-1})$,
414676. $(a=10^{-2})$, $1.57088 \cdot 10^8$ $(a=10^{-3})$, $4.87246 \cdot 10^{10}$
$(a=10^{-4})$,
$1.30774 \cdot 10^{13}$ 
$(a=10^{-5})$, $8.57852 \cdot 10^{27}$ $(a=0)$ (all seven
based on the first 
1,100,000 points of a scrambled Faure sequence).

Let us note that in terms of the Bures metric --- identically one-fourth
of the statistical distinguishability (SD) metric --- the pure state
[rank 1]
boundary of the $4 \times 4$ density matrices, {\it both} separable and
nonseparable, is known to have 
 volume $\frac{\pi^3}{6} \approx 5.16771$. (This is equal
to the volume of a 6-dimensional ball of radius 1 and to the volume
of a 3-dimensional complex projective space \cite[sec. IV.C]{hans1}.)
The 
14-dimensional \cite{bloore} submanifold 
 of $4 \times 4$ density matrices 
 of rank 3 has Bures volume $\frac{\pi^7}{4324320}
\approx 6.98444 \cdot 10^{-4}$. Multiplying by $2^{14}$, we obtain the SD
counterpart to this of 
\begin{equation} \label{dc}
B_{SD}^{s+n} = \frac{512  \pi^7}{135135} \approx 11.4433.
\end{equation}
This is {\it twice} (a ``double-covering'') the $(n-1)$-content (surface area 
or hyperarea) of the unit sphere in
$n=15$ dimensions.

\subsection{Further analysis}
Subsequently, 
we joined all the monotone metrics of interest into a single joint
analysis, using an independent Faure-Tezuka sequence of points in the
14-dimensional hypercube. Up to this point in time, we have generated
thirty-five million points (Table~\ref{tab:bounds}).
\begin{table}
\caption{\label{tab:bounds}Estimates of 
$14$-dimensional boundary volumes 
based on (four times) the Bures, Grosse-Krattenthaler-Slater, Wigner-Yanase, Average, Kubo-Mori and Noninformative metrics, using a scrambled Faure-Tezuka sequence composed of thirty-five million points.}
\begin{ruledtabular}
\begin{tabular}{rccc}
metric & $B^s$ & $B^{s+n}$ & $\beta$ \\
\hline
Bures & 0.456593 & 11.4443  & 0.584072 \\
GKS & ---& --- & 1.45204 \\
WY & 3203.81 & 17576.3 & 6606.58  \\
Avg & 6.60067 & 246.716 & 6.20592  \\
KM & --- &---  & 19.6215 \\
NI & --- & --- & 4333.36  \\
\end{tabular}
\end{ruledtabular}
\end{table}
For each of these points, we sought values of the fifteenth coordinate --- $\theta_{3}$ --- for which the partial transpose of the corresponding
$4 \times 4$ density matrix had zero determinant. For
24,038,658 of the points at least one feasible value of $\theta_{3}$
was found. The even-numbered solutions strongly dominated the 
odd-numbered
 solutions. (This may pertain to the fact that in this series of analyses,
we had used --- adjusting accordingly --- the range $[0,\pi]$, rather than $[0,\frac{\pi}{2}]$, as in our
other analyses, for $\theta_{1},\theta_{2}$ and $\theta_{3}$.) 
There were 74 points with one solution, 30 with three, and
11 with five, while there were 2,553,168  with two, 21,312,933 with four,
172,429 with six and 3 with eight.

\begin{figure}
\includegraphics{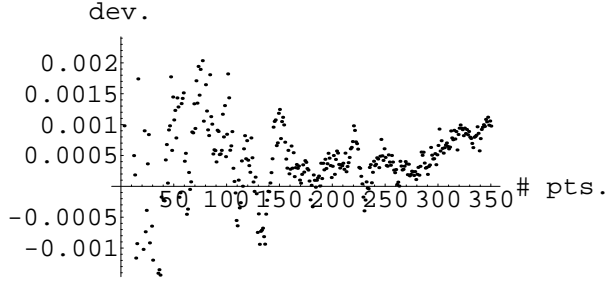}
\caption{\label{fig:sepplusnonsep}Deviations of the estimated value of
$B^{s+n}_{SD}$ from the {\it known} value of $\frac{512 \pi^7}{135135}
\approx 11.4433$, as the number of
points in a certain scrambled Faure-Tezuka 
sequence increases from one hundred thousand to thirty-five million}
\end{figure}

\begin{figure}
\includegraphics{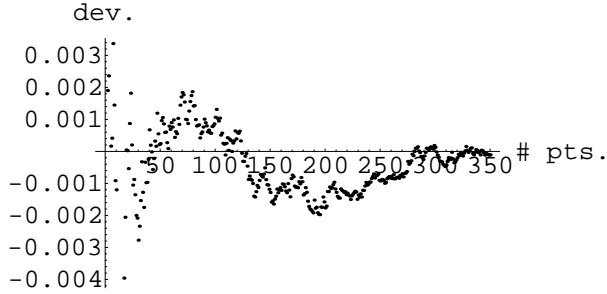}
\caption{\label{fig:sepboundary}Deviations of the 
cumulative estimates of $B^s_{SD}$
from the conjectured value of $\frac{43
 \sigma_{Ag}}{39} 
\approx 0.456697$. The number of points are recorded 
in steps of 100,000.}
\end{figure}

In Fig.~\ref{fig:sepplusnonsep}, we show the cumulative approximations (in steps  of one hundred thousand points) to
the known value (\ref{dc}) 
of $B_{SD}^{s+n}$.
We conjecture (Fig.~\ref{fig:sepboundary}) that 
the component of $B^{s+n}_{SD}$ consisting of separable states 
\cite{shidu}, that is $B^s_{SD}$, has the value $\frac{43  \sigma_{Ag}}{39}
\approx 0.456697$.
 The  concomitant estimate of the SD/Bures 
probability of separability 
of such rank-three states would then be
\begin{equation}
\Pi^s_{SD/Bures} = \frac{B^s_{SD}}{B^{s+n}_{SD}} = \frac{297297 
 \sigma_{Ag}}{1024  \pi^7} \approx 0.0398167 .
\end{equation}
This is considerably less than the
general probability of separability [of, generically, rank-{\it four} states], 
conjectured in formula (\ref{eqnew}) to be 0.0733389. The ratio of these two 
 probabilities
is $\frac{14157   \pi}{81920} \approx 0.542194$.

\begin{figure}
\includegraphics{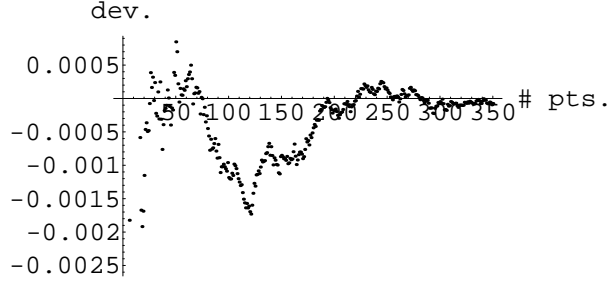}
\caption{\label{fig:rankfour}Deviations from the conjectured value of
$\beta_{SD} = \frac{55 \sigma_{Ag}}{39} \approx 0.584147$ of the cumulative estimates of the 
SD volume of that part of the 14-dimensional boundary of
the separable states consisting of nondegenerate states. The number of points
are recorded in steps of 100,000.}
\end{figure}

\begin{figure}
\includegraphics{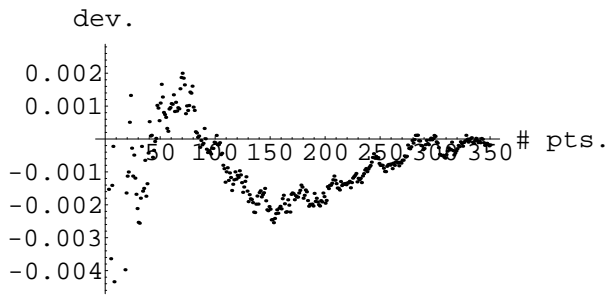}
\caption{\label{fig:total}Deviations from the conjectured value of
$B^s_{SD}+ \beta_{SD} = \frac{98 \sigma_{Ag}}{39}
\approx 1.04084$ of the SD-volume of
the {\it total} boundary of separable states --- composed of nondegenerate 
(rank-four)
and degenerate (rank-three) 
 $4 \times 4$ density matrices. The number of points 
of the Faure-Tezuka sequence are
recorded in steps of 100,000.}
\end{figure}

\begin{figure}
\includegraphics{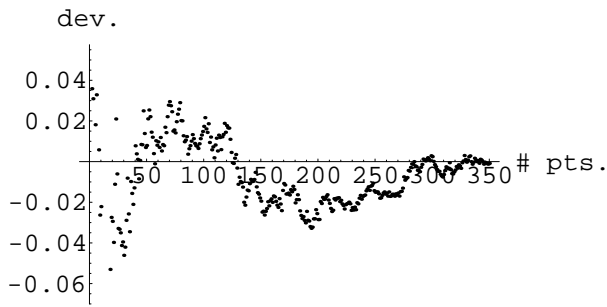}
\caption{\label{fig:avg}Deviations from the conjectured value of $\frac{255
 \sigma_{Ag}}{16} \approx 6.60153$
 of the cumulative estimates of $B^s_{\tilde{avg}}$.}
\end{figure}

\begin{figure}
\includegraphics{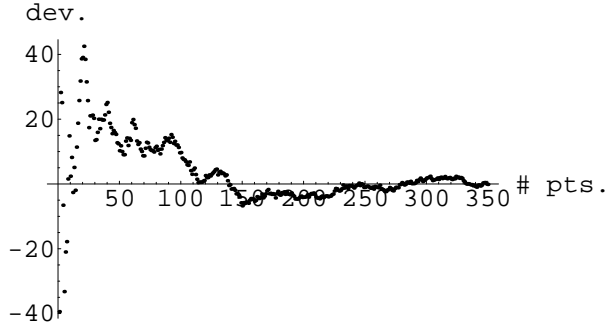}
\caption{\label{fig:WY1}Deviations from the conjectured value
of $7735 \sigma_{Ag} \approx 3203.94$ of the cumulative
estimates of $B^s_{\tilde{WY}}$}
\end{figure}

\begin{figure}
\includegraphics{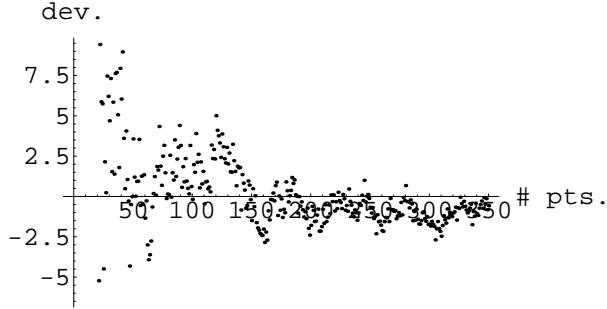}
\caption{\label{fig:WY}Deviations from the conjectured value of
$B^{s+n}_{\tilde{WY}}$ of
$\frac{262144 \pi^7}{45045} = 2^9 \cdot 3 B^{s+n}_{SD} \approx 
17576.9$.}
\end{figure}

\begin{figure}
\includegraphics{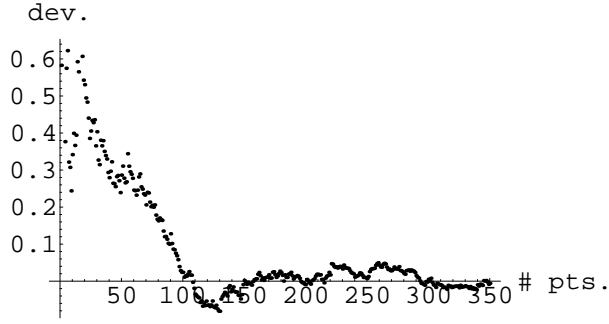}
\caption{\label{fig:KMinterior}Deviations from the conjectured value of
$\beta_{\tilde{KM}} = \frac{616  \sigma_{Ag}}{13} \approx 19.6274$}
\end{figure}

\begin{figure}
\includegraphics{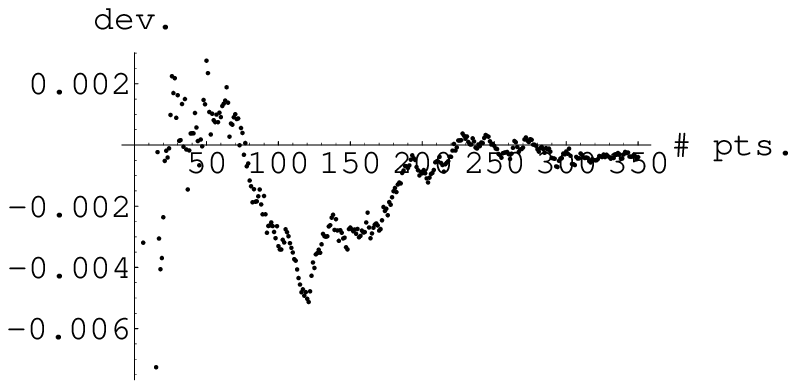}
\caption{\label{fig:GKSinterior}Deviations from the conjectured value
of $\beta_{\tilde{GKS}} = \frac{270 \sigma_{Ag}}{77} \approx 1.45244$}
\end{figure}

In Fig.~\ref{fig:rankfour}, we show the cumulative approximations to
a conjectured value of $\beta_{SD} =\frac{55 \sigma_{Ag}}{39} 
 \approx 0.584147$ for the 14-dimensional SD boundary
of separable two-qubit states composed of generically rank-{\it four} states.
(The test for membership in this class is that the determinant of the
partial transpose of the corresponding $4 \times 4$ density matrix be
zero.)
Thus, we have an implied conjecture that the 14-dimensional boundary
of separable $4 \times 4$ density matrices has {\it total} SD-volume
of $\beta_{SD}+B^s_{SD}=  \frac{98 \sigma_{Ag}}{39} 
\approx 1.04084$.
The fit of our cumulative estimates to this conjecture is shown in
Fig.~\ref{fig:total}.
Numerical evidence (Fig.~\ref{fig:avg}) also possibly suggests
 that $B^s_{\tilde{avg}} =\frac{255 \sigma_{Ag}}{16} \approx 6.60153$;
 that (Fig.~\ref{fig:WY1}) $B^s_{\tilde{WY}} = 7735 \sigma_{Ag} \approx 3203.94$; that (Fig.~\ref{fig:WY}) 
$B^{s+n}_{\tilde{WY}}= 2^9 \cdot 3 B^{s+n}_{SD}
\approx 17576.9$; that (Fig.~\ref{fig:KMinterior})
$\beta_{\tilde{KM}} =\frac{616 \sigma_{Ag}}{13}  \approx 19.6274$;
 and that (Fig.~\ref{fig:GKSinterior}), $\beta_{\tilde{GKS}}=\frac{270 \sigma_{ag}}{77} \approx 1.45244$.
So, surprisingly, the probability of separability of a rank-three state
appears to be much higher, that is, 
$\frac{348423075 \sigma_{Ag}}{262144  \pi^7} 
= \frac{3^2 \cdot 5^2 \cdot 7^2 \cdot 11 \cdot 13^2 \cdot 17  \sigma_{Ag}}{2^{18}  \pi^7} 
\approx 0.182281$ using the Wigner-Yanase  metric (but not the average 
metric, for which we have a sample estimate of 0.0267541 and a conjecture of
$\frac{10729125 \sigma_{Ag}}{54992 \pi^7} \approx 0.0267572$) than with the 
Bures or SD metric, in strong 
contrast to the rank-four case examined earlier (secs.~\ref{main} and \ref{Auxiliary}).

As is apparent, there are no estimates reported in Table~\ref{tab:bounds}
 for $B^s$ 
and $B^{s+n}$ for the GKS, KM and NI metrics. In retrospect, we could have
included the GKS metric (taking the limit of the corresponding volume element
as one of the eigenvalues of the associated $4 \times 4$ 
density matrix approaches zero),
 but the volume elements for the other two were
found to diverge on the rank-three states, so they could not have
been included.

\subsection{Levy-Gromov Isoperimetric Inequality}
The scalar curvature of the Bures metric for the $4 \times 4$ 
density matrices is bounded below by 570 \cite[Cor. 3]{ditt2}.
 However, application
of the Levy-Gromov Isoperimetric Inequality \cite[App. C]{gromov}
 requires a lower
bound of $N$ on $Ricci(Y,Y)$, where $Ricci$ is the Ricci tensor
and $Y$ runs over all unit
tangent vectors, for closed $(N+1)$-dimensional manifolds,
We did not immediately 
know if this condition is satisfied or not (given that one
can not
apparently ``control'' the Ricci curvature in terms of the
[bounded] scalar curvature), but we have found
that the inequality is violated, and that the condition is {\it not}
satisfied in the case before us.

To reach this conclusion, we first took the parameter $\alpha$ 
(strictly following the notation in \cite{gromov}) to be
$P_{SD/Bures} = \frac{V^s_{SD}}{V^{s+n}_{SD}} \approx 0.0736881$, according
to our conjecture (\ref{fundamental}) above. Then, 
the function 
$s(\alpha)$ is the
14-dimensional volume of the boundary sphere $\partial B_{\alpha}$,
where the volume of the ball $B_{\alpha}$ 
itself is equal to $\alpha$ $vol(S^{15})$,
and $S^{15}$ is the standard 15-dimensional sphere.
Further, 
the function 
$Is_{15}(\alpha)$ is the ratio of $s(\alpha) \approx 0.499459$ to $vol(S^{15}) = \frac{256 \pi^7}{2027025} \approx 0.381443$.
The Levy-Gromov Inequality then asserts that
$Is_{15}(\alpha)$, which here equals 
 1.30939, must be less than a certain ratio, which in our
case would --- according to our conjectures and known values --- be
\begin{equation}
 \frac{B^s_{Bures} + \beta_{Bures}}{V^{s+n}_{Bures}} = 
\frac{2^{-14} (B^s_{SD}+ \beta_{SD})}{2^{-15} V^{s+n}_{SD}} \approx 1.10573.
\end{equation}
So, the indicated inequality is violated.

At this point, we applied formula (7a) of \cite{ditt2}, giving the
Ricci tensor based on the Bures metric 
for {\it diagonal} density  matrices, 
\begin{equation} \label{dittRicci}
Ricci(Y,Z) = 3 \Sigma_{\mu,\nu,\eta}\frac{Y_{\nu \mu} \rho_{\eta} Z_{\mu \nu}}{(\rho_{\mu} +\rho_{\nu}) (\rho_{\mu}+\rho_{\eta}) (\rho_{\nu}+\rho_{\eta})}
-\frac{3}{2} \Sigma_{\mu,\nu}
\frac{Y_{\mu \mu} Z_{\nu \nu}}{(\rho_{\mu}+\rho_{\nu})^2},
\end{equation}
where $Y$ and $Z$ are tangent vectors (traceless Hermitian matrices).
We, in fact found, using 
numerical simulations, violations of the lower bound of $N$ for $(N-1)$-dimensional manifolds on the Ricci tensor required by the Levy-Gromov Inequality, 
with $N=14$ in our case. The lowest value we were able to achieve in 
a series of 
 simulations was 3.45666, so no {\it negative} values were recorded.
(The upper value appeared to be unbounded. We also applied the formula
(\ref{dittRicci}) to the 8-dimensional convex set of $3 \times 3$ density
matrices and found, through Monte Carlo 
simulations, a value of the Ricci curvature
as low as 3.00332, so it appears conjecturable that 3 is the actual lower
bound.)
Thus, our evidence here indicates that the inequality is 
not satisfied, apparently since 
 all the conditions for its application have not been met.
(Using the ansatz elaborated upon 
earler in sec.~\ref{MCF}, we also found
numerically 
for the Average and WY 
monotone metrics that the lower bound of $N=14$ on the Ricci tensor
was violated.)
\subsubsection{Area/Volume ratios}
Let us also note here (cf. \cite[sec. 6]{hans2}),
in terms of the known values  \cite{hans1} 
and our conjectures, that for the separable
{\it plus} nonseparable states, the ratio of the SD
 14-dimensional hyperarea to the SD 15-dimensional volume, is
\begin{equation}
\frac{B^{s+n}_{SD}}{V^{s+n}_{SD}} = \frac{8192}{429 \pi} \approx 6.07831
\end{equation}
while for only the separable states, it is
\begin{equation}
\frac{B^s_{SD}+\beta_{SD}}{V^s_{SD}} = \frac{98}{13} \approx 7.53846.
\end{equation}
\section{Concluding Remarks} \label{Concluding}
Needless to say, to the extent
 any of the conjectures above are, in fact, valid ones,
their remains the apparently formidable 
task of finding formal/rigorous proofs.

A direct/naive ``brute force''
strategy of {\it symbolically} integrating the volume
elements of the various monotone metrics over the {\it 15}-dimensional
convex sets of separable and all two-qubit states --- while successful
for lower-dimensional scenarios \cite{slaterC} --- appears to be
quite impractical computationally-speaking. It seems that one would have to deal
with multiple ranges of integration given by fourth-degree polynomials.
One might speculate that the integration of the (product) measures for the volume elements of  monotone metrics over the twelve parameterizing 
Euler angles would yield a result simply proportional to $\sigma_{Ag}$,
common to all monotone metrics, and that additional distinguishing factors
 would appear from integrating over the final
three variables ($\theta_{1},\theta_{2},\theta_{3}$ in the notation of
\cite{slaterqip}) parameterizing the eigenvalues of the $4 \times 4$ density
matrices.

Perhaps, in this regard, the work of Sommers and \.Zyczkowski
\cite{hans1} --- which they view ``as a contribution to the theory of
random matrices'' --- in constructing a  general formula for $V^{s+n}_{Bures}$ for $N$-level systems,
is extendible to (monotone) 
metrics {\it other} than the Bures. The volume 
$V^{s+n}_{SD}$ is known, and we have
indicated our conjectures that $V^{s+n}_{\tilde{KM}} =  
64 V^{s+n}_{SD}, V^{s+n}_{\tilde{avg}} = \frac{25 \pi^8}{8448}, 
V^{s+n}_{\tilde{GKS}}= \frac{\pi^8}{1750}$
 and $V^{s+n}_{\tilde{max}} =\infty$, 
but we have no similar conjecture, at the present,
for $V^{s+n}_{\tilde{WY}}$.
(The Wigner-Yanase metric corresponds
to a space of {\it constant curvature} \cite{gi}.)
But there appears to be no ``hint'' in the literature as to how one might 
{\it formally} derive simply the {\it separable} --- as
opposed to separable {\it plus} nonseparable --- volumes for any of the 
monotone metrics.

In this study, we have conjectured that the volumes of separable two-qubit
states is, as measured in terms of several monotone metrics 
of interest, simple multiples
of the silver mean ($\sigma_{ag}$). It is interesting to point out, it seems,
that in certain (``phyllotactic'')
 models of the arrangements of (rose) petals, the positions
of the petals (in fractions of a full turn) are given by the fractional
parts of simple multiples of the {\it golden} ratio \cite[p. 113]{livio}
\cite[pp. 122-123]{guy} (cf. \cite[p. 137]{mike}).
(Of course, it remains possible that we have been somewhat ``overeager''
here to find multiple roles for $\sigma_{Ag}$. In this regard, a sceptically-inclined reader
might point out that $2000 \sigma_{br} \approx 6605.55$ is  quite close
to our sample estimate of 6606.58 (Table~\ref{tab:bounds}) 
for $\beta_{\tilde{WY}}$ and
$2 \sigma_{br} \approx 6.60555$ approximates  our sample estimate of
6.60067 for $B^s_{\tilde{avg}}$. Here, $\sigma_{br}= \frac{3 +\sqrt{13}}{2}$ is
the ``bronze mean'' \cite{gumbs}.)

In this and other papers \cite{slaterqip,ZHSL,zycz2,slaterA,slaterC,hans2},
 attention has been focused on the matter 
of determining the volumes of quantum states in terms of various monotone
metrics. An even more considerable 
body of work concerned with  differential geometric properties
of the monotone metrics is devoted to issues of
the {\it scalar curvature} of monotone metrics \cite{gi,andai,ditty}.
For instance, it has been found that the scalar curvature of the $N \times N$
density matrices $D_{N}$  is, in terms of the Wigner-Yanase metric, 
$\frac{1}{4} (N^2-1)(N^2-2)$ \cite{gi}, while for the Bures metric, it is
24 for $N=2$ and for general $N$, no less than
 $\frac{1}{2}{(5 N^2-4) (N^2-1)}$, which value is assumed
 for the fully mixed state $\frac{1}{N} \bf{I}$ \cite{ditty}. 
(For the {\it infinite}-dimensional case of thermal squeezed states, 
Twamley \cite[eq. (30)]{twamley} has found the scalar curvature to be given
by $-\frac{8 (\cosh^2{\beta/4} + 12 \sinh^4{\beta/4})}{\cosh^2{\beta/2}}$, where the ``non-unitary'' parameter $\beta$ corresponds to the inverse temperature.)
It would certainly be of
interest to find linkages between these two 
interesting areas of investigation.
We note 
 that the {\it scalar curvature} determines the {\it asymptotic} behavior of
the {\it volume} of a Riemannian manifold \cite{gi}
 \cite[p. 55, Cor. 5.5 and ex. 3]{sakai}. Andai \cite[eq. (1)]{andairecent}
 has recently presented
a formula for the relation between the volume of a geodesic ball centered
at the fully mixed state and the scalar curvature there
(see also \cite[eq. (29)]{petzjpa}).

Our earlier conjecture (\ref{eq1}) --- in its unadjusted form --- as 
to the exact value of $V^s_{SD}$ had
suggested a similar-type conjecture for qubit-{\it qutrit} pairs
\cite{qq}. Now that we have found compelling numerical evidence to
reject (\ref{eq1}) (and replace it by (\ref{revised})), 
we obviously must be 
dubious as to the presumed validity of 
its qubit-qutrit analogue, but presently lack
any notion as to how to replace it. Additionally, our numerical
experience so far indicates that it would be extraordinarily difficult
to ``pinpoint'' (accurately estimate) the value of the volume of separable qubit-qutrit pairs,
since one would then be proceeding
 in a (more computationally demanding) 
much higher-dimensional ({\it 35}-d {\it v.} {\it 15}-d)
space, plus the size of the separable domain one would be
estimating would be much smaller relatively speaking (that is, relatively
fewer sampled $6 \times 6$ density matrices would be separable {\it vis-\'a-vis} the $4 \times 4$ case).

We summarize in Table~\ref{tab:table5} our present state of presumed knowledge
in regard to the various monotone metrics studied here. Of course, one would
aspire to find the {\it functionals} that map an operator monotone
function $f_{metric}(t)$ into  $V^s_{metric}, V^{s+n}_{metric}, B^s_{metric}, B^{s+n}_{metric}$ and $\beta_{metric}$. 
\begin{table}
\caption{\label{tab:table5} Conjectured  values (except for $V^{s+n}_{\tilde{Bures}}$ and $B^{s+n}_{\tilde{Bures}}$, 
which are {\it known}) of $V^s_{\tilde{metric}}$,
$V^{s+n}_{\tilde{metric}}$, $B^{s+n}_{\tilde{metric}}$,
$B^s_{\tilde{metric}}$ and $\beta_{\tilde{metric}}$ for 
(four times) various monotone metrics, listed in 
order of increasing volume size, together with the corresponding operator
monotone functions and Morozova-Chentsov functions. (For the various 
denominators,
we have the interesting 
prime decompositions: $5040 = 2^4 \cdot 3^2 \cdot 5 \cdot 7$; $1750 = 2 \cdot 5^3 \cdot 7$;
$8448 = 2^8 \cdot 3 \cdot 11$; $315 = 3^2 \cdot 5 \cdot 7$;
$135135 = 3^3 \cdot 5 \cdot 7 \cdot 11 \cdot 13$; $42075 = 3^2 \cdot 5^2 \cdot 11 \cdot 17$ and $45045 = 
3^2 \cdot 5 \cdot 7 \cdot 11 \cdot 13$. As pertains to numerators:
$512 = 2^9$; $7735 =  5  \cdot 7 \cdot  13 \cdot 17$; $262144 = 2^{18}$;  $15950  = 2 \cdot  5^2  \cdot 11 \cdot 29$; $255 = 3 \cdot 5 \cdot 17$; 
$495 = 3^3 \cdot 5 \cdot 11$; 
$270 = 2 \cdot 3^3 \cdot 5$; and $616 = 2^3 \cdot 7 \cdot 11$.)}
\begin{ruledtabular}
\begin{tabular}{rrrrrrrrr}
metric & $f(t)$ & $c(\rho_{\mu},\rho_{\nu})$ & $V^s$ & $V^{s+n}$ & $B^s$ & $B^{s+n}$ & $ \beta$  & $B^s+\beta$ \\
\hline
Bures & $\frac{1+t}{2}$ & $\frac{2}{\rho_{\mu}+\rho_{\nu}}$ & $\frac{\sigma_{Ag}}{3}$ & $\frac{\pi^8}{5040}$ & $\frac{43 \sigma_{Ag}}{39}$ & $\frac{512 \pi^7}{135135}$ & $ \frac{55 \sigma_{Ag}}{39}$ & $ \frac{98 \sigma_{Ag}}{39}$ \\
\hline
GKS & $\frac{t^{t/(t-1)}}{e}$ & $\frac{e (\frac{\rho_{\mu}}{\rho_{\nu}})^{\frac{\rho_{\mu}}{\rho_{\nu}-\rho_{\mu}}}}{\rho_{\nu}}$  & $\frac{4 \sigma_{Ag}}{5}$ & 
$\frac{\pi^8}{1750}$  & ?   & ? 
&  $\frac{270 \sigma_{Ag}}{77}$ & ?\\
\hline
WY & $\frac{(\sqrt{t}+1)^2}{4}$ & $\frac{4}{(\sqrt{\rho_{\mu}}+\sqrt{\rho_{\nu}})^2} $ & 
$\frac{7 \sigma_{Ag}}{4}$ & ? & $7735  \sigma_{Ag} $ &
 $\frac{262144 \pi^7}{45045}$ & $15950 \sigma_{Ag}$ & $23685 \sigma_{Ag}$\\
\hline
Avg & $\frac{1 + 6 t + t^2}{4 + 4 t}$ & $\frac{4 (\rho_{\mu}+\rho_{\nu})}{\rho_{\mu}^2 + 6 \rho_{\mu} \rho_{\nu} +\rho_{\nu}^2}$ &  $\frac{29 \sigma_{Ag}}{9}$
& $\frac{25 \pi^8}{8448}$ & $ \frac{255 \sigma_{Ag}}{16} $ & $\frac{3437 \pi^7}{42075}$  & $
15 \sigma_{Ag}$ & $\frac{495 \sigma_{Ag}}{16}$ \\
\hline
KM & $\frac{t-1}{\log{t}}$ & $\frac{\log{\frac{\rho_{\mu}}{\rho_{\nu}}}}{\rho_{\mu}-\rho_{\nu}}$ & $10 \sigma_{Ag}$ & $\frac{4 \pi^8}{315}$ & $\infty$ & $\infty$ & $\frac{616 \sigma_{Ag}}{13}$ & $\infty$\\
\hline
NI & $\frac{2 (t-1)^2}{(1+t) (\log{t})^2}$ & $\frac{(\rho_{\mu}+\rho_{\nu}) {\log^2{\frac{\rho_{\mu}}{\rho_{\nu}}}}}{2 (\rho_{\mu}-\rho_{\nu})^2}$ & ? & ? & $\infty$ & $\infty$
 & ? & $\infty$ \\ 
\hline
maximal & $\frac{2 t}{1+t}$ & $ \frac{\rho_{\mu}+\rho_{\nu}}{2 \rho_{\mu} \rho_{\nu}}$ & 
$\infty$ & $\infty$ & $\infty$ & $\infty$ & $\infty$ & $\infty$ \\
\end{tabular}
\end{ruledtabular}
\end{table}

\begin{acknowledgments}
I wish to express gratitude to the Kavli Institute for Theoretical
Physics for computational support in this research, to Giray \"Okten
for making available his MATHEMATICA programs for computing scrambled
Faure-Tezuka and Halton sequences, to A. Gherghetta for
correspondence regarding the definition of the silver mean, as well as to T. Isola  and P. Psaros. Also, K. \.Zyczkowski was very helpful in reviewing an
earlier draft. 
An anonymous referee, as well, made several valuable comments.

\end{acknowledgments}

\bibliography{Silver4d}

\begin{thebibliography}{71}
\expandafter\ifx\csname natexlab\endcsname\relax\def\natexlab#1{#1}\fi
\expandafter\ifx\csname bibnamefont\endcsname\relax
  \def\bibnamefont#1{#1}\fi
\expandafter\ifx\csname bibfnamefont\endcsname\relax
  \def\bibfnamefont#1{#1}\fi
\expandafter\ifx\csname citenamefont\endcsname\relax
  \def\citenamefont#1{#1}\fi
\expandafter\ifx\csname url\endcsname\relax
  \def\url#1{\texttt{#1}}\fi
\expandafter\ifx\csname urlprefix\endcsname\relax\def\urlprefix{URL }\fi
\providecommand{\bibinfo}[2]{#2}
\providecommand{\eprint}[2][]{\url{#2}}

\bibitem[{\citenamefont{Mkrtchian and Chaltykyan}(1987)}]{vanik}
\bibinfo{author}{\bibfnamefont{V.~E.} \bibnamefont{Mkrtchian}}
  \bibnamefont{and} \bibinfo{author}{\bibfnamefont{V.~O.}
  \bibnamefont{Chaltykyan}}, \bibinfo{journal}{Opt. Commun.}
  \textbf{\bibinfo{volume}{63}}, \bibinfo{pages}{239} (\bibinfo{year}{1987}).

\bibitem[{\citenamefont{Fano}(1983)}]{fano}
\bibinfo{author}{\bibfnamefont{U.}~\bibnamefont{Fano}}, \bibinfo{journal}{Rev.
  Mod. Phys.} \textbf{\bibinfo{volume}{55}}, \bibinfo{pages}{855}
  (\bibinfo{year}{1983}).

\bibitem[{\citenamefont{Braunstein and Caves}(1994)}]{sam}
\bibinfo{author}{\bibfnamefont{S.~L.} \bibnamefont{Braunstein}}
  \bibnamefont{and} \bibinfo{author}{\bibfnamefont{C.~M.} \bibnamefont{Caves}},
  \bibinfo{journal}{Phys. Rev. Lett.} \textbf{\bibinfo{volume}{72}},
  \bibinfo{pages}{3439} (\bibinfo{year}{1994}).

\bibitem[{\citenamefont{Slater}(2002)}]{slaterqip}
\bibinfo{author}{\bibfnamefont{P.~B.} \bibnamefont{Slater}},
  \bibinfo{journal}{Quant. Info. Proc.} \textbf{\bibinfo{volume}{1}},
  \bibinfo{pages}{397} (\bibinfo{year}{2002}).

\bibitem[{\citenamefont{{\.Z}yczkowski
  et~al.}(1998)\citenamefont{{\.Z}yczkowski, Horodecki, Sanpera, and
  Lewenstein}}]{ZHSL}
\bibinfo{author}{\bibfnamefont{K.}~\bibnamefont{{\.Z}yczkowski}},
  \bibinfo{author}{\bibfnamefont{P.}~\bibnamefont{Horodecki}},
  \bibinfo{author}{\bibfnamefont{A.}~\bibnamefont{Sanpera}}, \bibnamefont{and}
  \bibinfo{author}{\bibfnamefont{M.}~\bibnamefont{Lewenstein}},
  \bibinfo{journal}{Phys. Rev. A} \textbf{\bibinfo{volume}{58}},
  \bibinfo{pages}{883} (\bibinfo{year}{1998}).

\bibitem[{\citenamefont{{\.Z}yczkowski}(1999)}]{zycz2}
\bibinfo{author}{\bibfnamefont{K.}~\bibnamefont{{\.Z}yczkowski}},
  \bibinfo{journal}{Phys. Rev. A} \textbf{\bibinfo{volume}{60}},
  \bibinfo{pages}{3496} (\bibinfo{year}{1999}).

\bibitem[{\citenamefont{Slater}(1999{\natexlab{a}})}]{slaterA}
\bibinfo{author}{\bibfnamefont{P.~B.} \bibnamefont{Slater}},
  \bibinfo{journal}{J. Phys. A} \textbf{\bibinfo{volume}{32}},
  \bibinfo{pages}{5261} (\bibinfo{year}{1999}{\natexlab{a}}).

\bibitem[{\citenamefont{Slater}(2000)}]{slaterC}
\bibinfo{author}{\bibfnamefont{P.~B.} \bibnamefont{Slater}},
  \bibinfo{journal}{Euro. Phys. J. B} \textbf{\bibinfo{volume}{17}},
  \bibinfo{pages}{471} (\bibinfo{year}{2000}).

\bibitem[{\citenamefont{Werner}(1989)}]{werner}
\bibinfo{author}{\bibfnamefont{R.~F.} \bibnamefont{Werner}},
  \bibinfo{journal}{Phys. Rev. A} \textbf{\bibinfo{volume}{40}},
  \bibinfo{pages}{4277} (\bibinfo{year}{1989}).

\bibitem[{\citenamefont{Peres}(1996)}]{asher}
\bibinfo{author}{\bibfnamefont{A.}~\bibnamefont{Peres}},
  \bibinfo{journal}{Phys. Rev. Lett.} \textbf{\bibinfo{volume}{77}},
  \bibinfo{pages}{1413} (\bibinfo{year}{1996}).

\bibitem[{\citenamefont{Horodecki et~al.}(1996)\citenamefont{Horodecki,
  Horodecki, and Horodecki}}]{michal}
\bibinfo{author}{\bibfnamefont{M.}~\bibnamefont{Horodecki}},
  \bibinfo{author}{\bibfnamefont{P.}~\bibnamefont{Horodecki}},
  \bibnamefont{and}
  \bibinfo{author}{\bibfnamefont{R.}~\bibnamefont{Horodecki}},
  \bibinfo{journal}{Phys. Lett. A} \textbf{\bibinfo{volume}{223}},
  \bibinfo{pages}{1} (\bibinfo{year}{1996}).

\bibitem[{\citenamefont{Slater}(2003)}]{qq}
\bibinfo{author}{\bibfnamefont{P.~B.} \bibnamefont{Slater}},
  \bibinfo{journal}{J. Opt. B} \textbf{\bibinfo{volume}{5}},
  \bibinfo{pages}{S691} (\bibinfo{year}{2003}).

\bibitem[{\citenamefont{F.~Verstraete and Moor}(2001)}]{ver}
\bibinfo{author}{\bibfnamefont{J.}~\bibnamefont{F.~Verstraete}}
  \bibnamefont{and} \bibinfo{author}{\bibfnamefont{B.~D.} \bibnamefont{Moor}},
  \bibinfo{journal}{Phys. Rev. A} \textbf{\bibinfo{volume}{64}},
  \bibinfo{pages}{010101} (\bibinfo{year}{2001}).

\bibitem[{\citenamefont{Sommers and {\.Z}yczkowski}(2003)}]{hans1}
\bibinfo{author}{\bibfnamefont{H.-J.} \bibnamefont{Sommers}} \bibnamefont{and}
  \bibinfo{author}{\bibfnamefont{K.}~\bibnamefont{{\.Z}yczkowski}},
  \bibinfo{journal}{J. Phys. A} \textbf{\bibinfo{volume}{36}},
  \bibinfo{pages}{10883} (\bibinfo{year}{2003}).

\bibitem[{\citenamefont{Slater}(1999{\natexlab{b}})}]{slaterhall}
\bibinfo{author}{\bibfnamefont{P.~B.} \bibnamefont{Slater}},
  \bibinfo{journal}{J. Phys. A} \textbf{\bibinfo{volume}{32}},
  \bibinfo{pages}{8231} (\bibinfo{year}{1999}{\natexlab{b}}).

\bibitem[{\citenamefont{{\.Z}yczkowski and Sommers}(2003)}]{hans2}
\bibinfo{author}{\bibfnamefont{K.}~\bibnamefont{{\.Z}yczkowski}}
  \bibnamefont{and} \bibinfo{author}{\bibfnamefont{H.-J.}
  \bibnamefont{Sommers}}, \bibinfo{journal}{J. Phys. A}
  \textbf{\bibinfo{volume}{36}}, \bibinfo{pages}{10115} (\bibinfo{year}{2003}).

\bibitem[{\citenamefont{Tilma et~al.}(2002)\citenamefont{Tilma, Byrd, and
  Sudarshan}}]{tbs}
\bibinfo{author}{\bibfnamefont{T.}~\bibnamefont{Tilma}},
  \bibinfo{author}{\bibfnamefont{M.}~\bibnamefont{Byrd}}, \bibnamefont{and}
  \bibinfo{author}{\bibfnamefont{E.~C.~G.} \bibnamefont{Sudarshan}},
  \bibinfo{journal}{J. Phys. A} \textbf{\bibinfo{volume}{35}},
  \bibinfo{pages}{10445} (\bibinfo{year}{2002}).

\bibitem[{\citenamefont{Lyness}(1965)}]{lyness}
\bibinfo{author}{\bibfnamefont{J.~N.} \bibnamefont{Lyness}},
  \bibinfo{journal}{Math. Comput.} \textbf{\bibinfo{volume}{19}},
  \bibinfo{pages}{394} (\bibinfo{year}{1965}).

\bibitem[{\citenamefont{{\"O}kten}(1999)}]{giray1}
\bibinfo{author}{\bibfnamefont{G.}~\bibnamefont{{\"O}kten}},
  \bibinfo{journal}{MATHEMATICA in Educ. Res.} \textbf{\bibinfo{volume}{8}},
  \bibinfo{pages}{52} (\bibinfo{year}{1999}).

\bibitem[{\citenamefont{Petz and Sud\mbox{\'a}r}(1996)}]{petz1}
\bibinfo{author}{\bibfnamefont{D.}~\bibnamefont{Petz}} \bibnamefont{and}
  \bibinfo{author}{\bibfnamefont{C.}~\bibnamefont{Sud\mbox{\'a}r}},
  \bibinfo{journal}{J. Math. Phys.} \textbf{\bibinfo{volume}{37}},
  \bibinfo{pages}{2662} (\bibinfo{year}{1996}).

\bibitem[{\citenamefont{Petz}(1996)}]{petz2}
\bibinfo{author}{\bibfnamefont{D.}~\bibnamefont{Petz}}, \bibinfo{journal}{Lin.
  Alg. Applics.} \textbf{\bibinfo{volume}{244}}, \bibinfo{pages}{81}
  (\bibinfo{year}{1996}).

\bibitem[{\citenamefont{Lesniewski and Ruskai}(1999)}]{lesniewski}
\bibinfo{author}{\bibfnamefont{A.}~\bibnamefont{Lesniewski}} \bibnamefont{and}
  \bibinfo{author}{\bibfnamefont{M.~B.} \bibnamefont{Ruskai}},
  \bibinfo{journal}{J. Math. Phys.} \textbf{\bibinfo{volume}{40}},
  \bibinfo{pages}{5702} (\bibinfo{year}{1999}).

\bibitem[{\citenamefont{Kass}(1997)}]{kass}
\bibinfo{author}{\bibfnamefont{R.~E.} \bibnamefont{Kass}},
  \emph{\bibinfo{title}{Geometrical Foundations of Asymptotic Inference}}
  (\bibinfo{publisher}{John Wiley}, \bibinfo{address}{New York},
  \bibinfo{year}{1997}).

\bibitem[{\citenamefont{H{\"u}bner}(1992)}]{hubner1}
\bibinfo{author}{\bibfnamefont{M.}~\bibnamefont{H{\"u}bner}},
  \bibinfo{journal}{Phys. Lett. A} \textbf{\bibinfo{volume}{63}},
  \bibinfo{pages}{239} (\bibinfo{year}{1992}).

\bibitem[{\citenamefont{H{\"u}bner}(179)}]{hubner2}
\bibinfo{author}{\bibfnamefont{M.}~\bibnamefont{H{\"u}bner}},
  \bibinfo{journal}{Phys. Lett. A} \textbf{\bibinfo{volume}{179}},
  \bibinfo{pages}{226} (\bibinfo{year}{179}).

\bibitem[{\citenamefont{Dittmann}(1999{\natexlab{a}})}]{ditt1}
\bibinfo{author}{\bibfnamefont{J.}~\bibnamefont{Dittmann}},
  \bibinfo{journal}{J. Phys. A} \textbf{\bibinfo{volume}{32}},
  \bibinfo{pages}{2663} (\bibinfo{year}{1999}{\natexlab{a}}).

\bibitem[{\citenamefont{Dittmann}(1999{\natexlab{b}})}]{ditt2}
\bibinfo{author}{\bibfnamefont{J.}~\bibnamefont{Dittmann}},
  \bibinfo{journal}{J. Geom. Phys.} \textbf{\bibinfo{volume}{31}},
  \bibinfo{pages}{16} (\bibinfo{year}{1999}{\natexlab{b}}).

\bibitem[{\citenamefont{Yuen and Lax}(1973)}]{yuenlax}
\bibinfo{author}{\bibfnamefont{H.~P.} \bibnamefont{Yuen}} \bibnamefont{and}
  \bibinfo{author}{\bibfnamefont{M.}~\bibnamefont{Lax}}, \bibinfo{journal}{IEEE
  Trans. Inform. Th.} \textbf{\bibinfo{volume}{19}}, \bibinfo{pages}{740}
  (\bibinfo{year}{1973}).

\bibitem[{\citenamefont{Hasegawa}(1997)}]{hasegawa}
\bibinfo{author}{\bibfnamefont{H.}~\bibnamefont{Hasegawa}},
  \bibinfo{journal}{Rep. Math. Phys.} \textbf{\bibinfo{volume}{39}},
  \bibinfo{pages}{49} (\bibinfo{year}{1997}).

\bibitem[{\citenamefont{Petz}(1994)}]{petz3}
\bibinfo{author}{\bibfnamefont{D.}~\bibnamefont{Petz}}, \bibinfo{journal}{J.
  Math, Phys.} \textbf{\bibinfo{volume}{35}}, \bibinfo{pages}{780}
  (\bibinfo{year}{1994}).

\bibitem[{\citenamefont{Michor et~al.}(2002)\citenamefont{Michor, Petz, and
  Andai}}]{michor}
\bibinfo{author}{\bibfnamefont{P.~W.} \bibnamefont{Michor}},
  \bibinfo{author}{\bibfnamefont{D.}~\bibnamefont{Petz}}, \bibnamefont{and}
  \bibinfo{author}{\bibfnamefont{A.}~\bibnamefont{Andai}},
  \bibinfo{journal}{Infin. Dimens. Anal. Quantum Probab. Relat. Top.}
  \textbf{\bibinfo{volume}{3}}, \bibinfo{pages}{199} (\bibinfo{year}{2002}).

\bibitem[{\citenamefont{Grasselli and Streater}(2001)}]{streater}
\bibinfo{author}{\bibfnamefont{M.~R.} \bibnamefont{Grasselli}}
  \bibnamefont{and} \bibinfo{author}{\bibfnamefont{R.~F.}
  \bibnamefont{Streater}}, \bibinfo{journal}{Infin. Dimens. Anal. Quantum
  Probab. Relat. Top.} \textbf{\bibinfo{volume}{4}}, \bibinfo{pages}{173}
  (\bibinfo{year}{2001}).

\bibitem[{\citenamefont{Faure and Tezuka}(2002)}]{tezuka}
\bibinfo{author}{\bibfnamefont{H.}~\bibnamefont{Faure}} \bibnamefont{and}
  \bibinfo{author}{\bibfnamefont{S.}~\bibnamefont{Tezuka}}, in
  \emph{\bibinfo{booktitle}{Monte Carlo and Quasi-Monte Carlo Methods 2000
  (Hong Kong)}}, edited by \bibinfo{editor}{\bibfnamefont{K.~T.}
  \bibnamefont{Tang}}, \bibinfo{editor}{\bibfnamefont{F.~J.}
  \bibnamefont{Hickernell}}, \bibnamefont{and}
  \bibinfo{editor}{\bibfnamefont{H.}~\bibnamefont{Niederreiter}}
  (\bibinfo{publisher}{Springer}, \bibinfo{address}{Berlin},
  \bibinfo{year}{2002}), p. \bibinfo{pages}{242}.

\bibitem[{\citenamefont{Call and Velleman}(1993)}]{call}
\bibinfo{author}{\bibfnamefont{G.}~\bibnamefont{Call}} \bibnamefont{and}
  \bibinfo{author}{\bibfnamefont{D.}~\bibnamefont{Velleman}},
  \bibinfo{journal}{Amer. Math. Monthly} \textbf{\bibinfo{volume}{100}},
  \bibinfo{pages}{372} (\bibinfo{year}{1993}).

\bibitem[{\citenamefont{Hong and Hickernell}(2003)}]{hong}
\bibinfo{author}{\bibfnamefont{H.~S.} \bibnamefont{Hong}} \bibnamefont{and}
  \bibinfo{author}{\bibfnamefont{F.~J.} \bibnamefont{Hickernell}},
  \bibinfo{journal}{ACM Trans. Math. Software} \textbf{\bibinfo{volume}{29}},
  \bibinfo{pages}{95} (\bibinfo{year}{2003}).

\bibitem[{\citenamefont{Mato{\u u}sek}(1999)}]{matou}
\bibinfo{author}{\bibfnamefont{J.}~\bibnamefont{Mato{\u u}sek}},
  \emph{\bibinfo{title}{Geometric Discrepancy: An Illustrated Guide}}
  (\bibinfo{publisher}{Springer}, \bibinfo{address}{Berlin},
  \bibinfo{year}{1999}).

\bibitem[{\citenamefont{Hall}(1998)}]{hall}
\bibinfo{author}{\bibfnamefont{M.~J.~W.} \bibnamefont{Hall}},
  \bibinfo{journal}{Phys. Lett. A} \textbf{\bibinfo{volume}{242}},
  \bibinfo{pages}{123} (\bibinfo{year}{1998}).

\bibitem[{\citenamefont{Christos and Gherghetta}(1991)}]{christos}
\bibinfo{author}{\bibfnamefont{G.~A.} \bibnamefont{Christos}} \bibnamefont{and}
  \bibinfo{author}{\bibfnamefont{T.}~\bibnamefont{Gherghetta}},
  \bibinfo{journal}{Phys. Rev. A} \textbf{\bibinfo{volume}{44}},
  \bibinfo{pages}{898} (\bibinfo{year}{1991}).

\bibitem[{\citenamefont{Abe and Rajagopal}(1999)}]{abe}
\bibinfo{author}{\bibfnamefont{S.}~\bibnamefont{Abe}} \bibnamefont{and}
  \bibinfo{author}{\bibfnamefont{A.~K.} \bibnamefont{Rajagopal}},
  \bibinfo{journal}{Phys. Rev. A} \textbf{\bibinfo{volume}{60}},
  \bibinfo{pages}{3461} (\bibinfo{year}{1999}).

\bibitem[{\citenamefont{Gill et~al.}(2002)\citenamefont{Gill, Weihs, Zeilinger,
  and {\.Z}ukowski}}]{gill}
\bibinfo{author}{\bibfnamefont{R.~D.} \bibnamefont{Gill}},
  \bibinfo{author}{\bibfnamefont{G.}~\bibnamefont{Weihs}},
  \bibinfo{author}{\bibfnamefont{A.}~\bibnamefont{Zeilinger}},
  \bibnamefont{and}
  \bibinfo{author}{\bibfnamefont{M.}~\bibnamefont{{\.Z}ukowski}},
  \bibinfo{journal}{Proc. Natl. Acad. Sci.} \textbf{\bibinfo{volume}{99}},
  \bibinfo{pages}{14632} (\bibinfo{year}{2002}).

\bibitem[{\citenamefont{de~Spinadel}(2002)}]{spinadel}
\bibinfo{author}{\bibfnamefont{V.~W.} \bibnamefont{de~Spinadel}},
  \bibinfo{journal}{Int. Math. J.} \textbf{\bibinfo{volume}{2}},
  \bibinfo{pages}{279} (\bibinfo{year}{2002}).

\bibitem[{\citenamefont{Gumbs}(1989)}]{gumbs}
\bibinfo{author}{\bibfnamefont{G.}~\bibnamefont{Gumbs}}, \bibinfo{journal}{J.
  Phys. A} \textbf{\bibinfo{volume}{22}}, \bibinfo{pages}{951}
  (\bibinfo{year}{1989}).

\bibitem[{\citenamefont{Kappraff}(2002)}]{kappraff}
\bibinfo{author}{\bibfnamefont{J.}~\bibnamefont{Kappraff}},
  \emph{\bibinfo{title}{Beyond Measure}} (\bibinfo{publisher}{World
  Scientific}, \bibinfo{address}{River Edge, NJ}, \bibinfo{year}{2002}).

\bibitem[{\citenamefont{Escudero and Garcia}(2001)}]{escudero}
\bibinfo{author}{\bibfnamefont{J.~G.} \bibnamefont{Escudero}} \bibnamefont{and}
  \bibinfo{author}{\bibfnamefont{J.~G.} \bibnamefont{Garcia}},
  \bibinfo{journal}{J. Phys. Soc. Japan.} \textbf{\bibinfo{volume}{70}},
  \bibinfo{pages}{3511} (\bibinfo{year}{2001}).

\bibitem[{\citenamefont{\mbox{El} Naschie}(1999)}]{carlosfriend}
\bibinfo{author}{\bibfnamefont{M.~S.} \bibnamefont{\mbox{El} Naschie}},
  \bibinfo{journal}{Chaos, Sol. Fract.} \textbf{\bibinfo{volume}{10}},
  \bibinfo{pages}{1303} (\bibinfo{year}{1999}).

\bibitem[{\citenamefont{Chuan and Yu}(2000)}]{chuan}
\bibinfo{author}{\bibfnamefont{W.-F.} \bibnamefont{Chuan}} \bibnamefont{and}
  \bibinfo{author}{\bibfnamefont{F.}~\bibnamefont{Yu}},
  \bibinfo{journal}{Fibonacci Quart.} \textbf{\bibinfo{volume}{38}},
  \bibinfo{pages}{425} (\bibinfo{year}{2000}).

\bibitem[{\citenamefont{Bavard and Pansu}(1986)}]{bayard}
\bibinfo{author}{\bibfnamefont{C.}~\bibnamefont{Bavard}} \bibnamefont{and}
  \bibinfo{author}{\bibfnamefont{P.}~\bibnamefont{Pansu}},
  \bibinfo{journal}{Ann. Sci. \'Ecole Norm. Sup.}
  \textbf{\bibinfo{volume}{19}}, \bibinfo{pages}{479} (\bibinfo{year}{1986}).

\bibitem[{\citenamefont{Bowditch}(1993)}]{bowditch}
\bibinfo{author}{\bibfnamefont{B.~H.} \bibnamefont{Bowditch}},
  \bibinfo{journal}{J. Austral. Math. Soc. Ser. A}
  \textbf{\bibinfo{volume}{55}}, \bibinfo{pages}{23} (\bibinfo{year}{1993}).

\bibitem[{\citenamefont{Bambah et~al.}(1986)\citenamefont{Bambah, Dumir, and
  Hans-Gill}}]{bambah}
\bibinfo{author}{\bibfnamefont{R.~P.} \bibnamefont{Bambah}},
  \bibinfo{author}{\bibfnamefont{V.~C.} \bibnamefont{Dumir}}, \bibnamefont{and}
  \bibinfo{author}{\bibfnamefont{R.~J.} \bibnamefont{Hans-Gill}},
  \bibinfo{journal}{Studia Sci. Math. Hungar.} \textbf{\bibinfo{volume}{21}},
  \bibinfo{pages}{135} (\bibinfo{year}{1986}).

\bibitem[{\citenamefont{Krattenthaler and Slater}(2000)}]{KS}
\bibinfo{author}{\bibfnamefont{C.}~\bibnamefont{Krattenthaler}}
  \bibnamefont{and} \bibinfo{author}{\bibfnamefont{P.~B.}
  \bibnamefont{Slater}}, \bibinfo{journal}{IEEE. Trans. Inform. Th.}
  \textbf{\bibinfo{volume}{46}}, \bibinfo{pages}{801} (\bibinfo{year}{2000}).

\bibitem[{\citenamefont{Slater}(2001)}]{gillmassar}
\bibinfo{author}{\bibfnamefont{P.~B.} \bibnamefont{Slater}},
  \bibinfo{journal}{J. Phys. A} \textbf{\bibinfo{volume}{34}},
  \bibinfo{pages}{7029} (\bibinfo{year}{2001}).

\bibitem[{\citenamefont{Gibilisco and Isola}(2003)}]{gi}
\bibinfo{author}{\bibfnamefont{P.}~\bibnamefont{Gibilisco}} \bibnamefont{and}
  \bibinfo{author}{\bibfnamefont{T.}~\bibnamefont{Isola}}, \bibinfo{journal}{J.
  Math. Phys.} \textbf{\bibinfo{volume}{44}}, \bibinfo{pages}{3752}
  (\bibinfo{year}{2003}).

\bibitem[{\citenamefont{Wigner and Yanase}(1963)}]{wy}
\bibinfo{author}{\bibfnamefont{E.}~\bibnamefont{Wigner}} \bibnamefont{and}
  \bibinfo{author}{\bibfnamefont{M.}~\bibnamefont{Yanase}},
  \bibinfo{journal}{Proc. Natl. Acad. Sci.} \textbf{\bibinfo{volume}{49}},
  \bibinfo{pages}{910} (\bibinfo{year}{1963}).

\bibitem[{\citenamefont{Luo}(2003)}]{luo}
\bibinfo{author}{\bibfnamefont{S.}~\bibnamefont{Luo}}, \bibinfo{journal}{Phys.
  Rev. Lett.} \textbf{\bibinfo{volume}{91}}, \bibinfo{pages}{180403}
  (\bibinfo{year}{2003}).

\bibitem[{\citenamefont{Slater}(1998)}]{slaterclarke}
\bibinfo{author}{\bibfnamefont{P.~B.} \bibnamefont{Slater}},
  \bibinfo{journal}{Phys. Lett. A} \textbf{\bibinfo{volume}{247}},
  \bibinfo{pages}{1} (\bibinfo{year}{1998}).

\bibitem[{\citenamefont{Braunstein et~al.}(1999)\citenamefont{Braunstein,
  Caves, Jozsa, Linden, Popescu, and Schack}}]{slb}
\bibinfo{author}{\bibfnamefont{S.~L.} \bibnamefont{Braunstein}},
  \bibinfo{author}{\bibfnamefont{C.~M.} \bibnamefont{Caves}},
  \bibinfo{author}{\bibfnamefont{R.}~\bibnamefont{Jozsa}},
  \bibinfo{author}{\bibfnamefont{N.}~\bibnamefont{Linden}},
  \bibinfo{author}{\bibfnamefont{S.}~\bibnamefont{Popescu}}, \bibnamefont{and}
  \bibinfo{author}{\bibfnamefont{R.}~\bibnamefont{Schack}},
  \bibinfo{journal}{Phys. Rev. Lett.} \textbf{\bibinfo{volume}{83}},
  \bibinfo{pages}{1054} (\bibinfo{year}{1999}).

\bibitem[{\citenamefont{Clifton and Halvorson}(2000)}]{ch}
\bibinfo{author}{\bibfnamefont{R.}~\bibnamefont{Clifton}} \bibnamefont{and}
  \bibinfo{author}{\bibfnamefont{H.}~\bibnamefont{Halvorson}},
  \bibinfo{journal}{Phys. Rev. A} \textbf{\bibinfo{volume}{61}},
  \bibinfo{pages}{0121018} (\bibinfo{year}{2000}).

\bibitem[{\citenamefont{Gurvits and Barnum}(2003)}]{gb}
\bibinfo{author}{\bibfnamefont{L.}~\bibnamefont{Gurvits}} \bibnamefont{and}
  \bibinfo{author}{\bibfnamefont{H.}~\bibnamefont{Barnum}},
  \bibinfo{journal}{Phys. Rev. A} \textbf{\bibinfo{volume}{68}},
  \bibinfo{pages}{042310} (\bibinfo{year}{2003}).

\bibitem[{\citenamefont{Szarek}()}]{szarek}
\bibinfo{author}{\bibfnamefont{S.}~\bibnamefont{Szarek}},
  \eprint{quant-ph/0310061}.

\bibitem[{\citenamefont{Bloore}(1976)}]{bloore}
\bibinfo{author}{\bibfnamefont{F.~J.} \bibnamefont{Bloore}},
  \bibinfo{journal}{J. Phys. A.} \textbf{\bibinfo{volume}{9}},
  \bibinfo{pages}{2059} (\bibinfo{year}{1976}).

\bibitem[{\citenamefont{Shi and Du}()}]{shidu}
\bibinfo{author}{\bibfnamefont{M.}~\bibnamefont{Shi}} \bibnamefont{and}
  \bibinfo{author}{\bibfnamefont{J.}~\bibnamefont{Du}},
  \eprint{quant-ph/0103016}.

\bibitem[{\citenamefont{Gromov}(1999)}]{gromov}
\bibinfo{author}{\bibfnamefont{M.}~\bibnamefont{Gromov}},
  \emph{\bibinfo{title}{Metric Structures for Riemannian and Non-Riemannian
  Spaces}} (\bibinfo{publisher}{Birkh{\"a}user}, \bibinfo{address}{Boston},
  \bibinfo{year}{1999}).

\bibitem[{\citenamefont{Livio}(2002)}]{livio}
\bibinfo{author}{\bibfnamefont{M.}~\bibnamefont{Livio}},
  \emph{\bibinfo{title}{The Golden Ratio}} (\bibinfo{publisher}{Broadway},
  \bibinfo{address}{New York}, \bibinfo{year}{2002}).

\bibitem[{\citenamefont{Conway and Guy}(1996)}]{guy}
\bibinfo{author}{\bibfnamefont{J.~H.} \bibnamefont{Conway}} \bibnamefont{and}
  \bibinfo{author}{\bibfnamefont{R.~K.} \bibnamefont{Guy}},
  \emph{\bibinfo{title}{The Book of Numbers}}
  (\bibinfo{publisher}{Springer-Verlag}, \bibinfo{address}{New York},
  \bibinfo{year}{1996}).

\bibitem[{\citenamefont{Freedman}(2003)}]{mike}
\bibinfo{author}{\bibfnamefont{M.~H.} \bibnamefont{Freedman}},
  \bibinfo{journal}{Comm. Math. Phys.} \textbf{\bibinfo{volume}{234}},
  \bibinfo{pages}{129} (\bibinfo{year}{2003}).

\bibitem[{\citenamefont{Andai}(2003)}]{andai}
\bibinfo{author}{\bibfnamefont{A.}~\bibnamefont{Andai}}, \bibinfo{journal}{J.
  Math. Phys.} \textbf{\bibinfo{volume}{44}}, \bibinfo{pages}{3676}
  (\bibinfo{year}{2003}).

\bibitem[{\citenamefont{Dittmann}(1999{\natexlab{c}})}]{ditty}
\bibinfo{author}{\bibfnamefont{J.}~\bibnamefont{Dittmann}},
  \bibinfo{journal}{J. Geom. Phys.} \textbf{\bibinfo{volume}{31}},
  \bibinfo{pages}{16} (\bibinfo{year}{1999}{\natexlab{c}}).

\bibitem[{\citenamefont{Twamley}(1996)}]{twamley}
\bibinfo{author}{\bibfnamefont{J.}~\bibnamefont{Twamley}}, \bibinfo{journal}{J.
  Phys. A} \textbf{\bibinfo{volume}{29}}, \bibinfo{pages}{3723}
  (\bibinfo{year}{1996}).

\bibitem[{\citenamefont{Sakai}(1996)}]{sakai}
\bibinfo{author}{\bibfnamefont{T.}~\bibnamefont{Sakai}},
  \emph{\bibinfo{title}{Riemannian Geometry}} (\bibinfo{publisher}{Amer. Math.
  Soc.}, \bibinfo{address}{Providence}, \bibinfo{year}{1996}).

\bibitem[{\citenamefont{Andai}()}]{andairecent}
\bibinfo{author}{\bibfnamefont{A.}~\bibnamefont{Andai}},
  \eprint{math-ph/0310064}.

\bibitem[{\citenamefont{Petz}(2002)}]{petzjpa}
\bibinfo{author}{\bibfnamefont{D.}~\bibnamefont{Petz}}, \bibinfo{journal}{J.
  Phys. A} \textbf{\bibinfo{volume}{35}}, \bibinfo{pages}{929}
  (\bibinfo{year}{2002}).

\end{thebibliography}

\end{document}